hep-th/9801171

\def\Ad            {\mathrm{Ad}}
\def\ala           {\alpha_\lambda}
\def\alap          {\alpha^+_\lambda}
\def\alam          {\alpha^-_\lambda}
\def\amu           {\alpha_\mu}

\def\amum          {\alpha^-_\mu}

\def\alab          {\alpha_{\lambdab}}
\def\AI            {{A(I)}}
\def\AIo           {{A(\Io)}}

\def\Aut           {\mathrm{Aut}}
\def\bala          {\overline{\alpha}_\lambda}
\def\balab         {\overline{\alpha}_{\lambdab}}
\def\bbC           {\mathbb{C}}
\def\bbM           {\mathbb{M}}
\def\bbN           {\mathbb{N}}

\def\bbZ           {\mathbb{Z}}
\def\be            {\begin{equation}}
\def\bearl         {\begin{array}{l}}
\def\bearll        {\begin{array}{ll}}
\def\bearlll       {\begin{array}{lll}}
\def\bearrl        {\begin{array}{rl}}
\def\bea           {\begin{eqnarray}}
\def\beaa          {\begin{eqnarray*}}
\def\bfe           {{\bf1}}

\newcommand\bproof {\noindent {\it Proof. }}

\def\can           {\gamma}
\def\cani          {\gamma^{-1}}
\def\canr          {\theta}

\def\cA            {\mathcal{A}}
\def\cC            {\mathcal{C}}

\def\cF            {\mathcal{F}}

\def\cH            {\mathcal{H}}
\def\cJ            {\mathcal{J}}
\def\cM            {\mathcal{M}}
\def\cN            {\mathcal{N}}

\def\cM            {\mathcal{M}}
\newcommand\co[1]  {\overline{{#1}}}
\def\cO            {\mathcal{O}}
\def\CA            {\cC_\cA}
\def\CAo           {\cC_\cA^{(0)}}

\def\cT            {\mathcal{T}}
\def\cV            {\mathcal{V}}
\def\cW            {\mathcal{W}}
\def\CA            {\cC_\cA}
\def\CAo           {\cC_\cA^{(0)}}
\def\CN            {\cC_\cN}

\def\DelAI         {\Delta_\cA(I)}
\def\DelAIo        {\Delta_\cA(\Io)}

\def\DelNIo        {\Delta_\cN(\Io)}
\def\DelMIo        {\Delta_\cM(\Io)}
\def\DelMIO        {\Delta_\cM^{(0)}(I)}
\def\DelMIoO       {\Delta_\cM^{(0)}(\Io)}

\def\DiffS         {\mathrm{Diff}(S^1)}

\def\dim           {\mathrm{dim}}
\newcommand\del[2] {\delta_{{#1},{#2}}}

\def\ee            {\end{equation}}
\def\eear          {\end{array}}
\def\eea           {\end{eqnarray}}
\def\eeaa          {\end{eqnarray*}}

\def\End           {\mathrm{End}}

\newcommand\eproof {\hspace*{\fill}\nolinebreak\hspace*{\fill}
                   {\sl Q.E.D.}
                   \par\vspace{3mm}}
\newcommand\eps[2] {\varepsilon({#1},{#2})}
\newcommand\epsm[2]{\varepsilon^-({#1},{#2})}
\newcommand\epsp[2]{\varepsilon^+({#1},{#2})}
\newcommand\epsmp[2]{\varepsilon^\mp({#1},{#2})}
\newcommand\epspm[2]{\varepsilon^\pm({#1},{#2})}

\newcommand\epsom[2]{\epsilon^-({#1},{#2})}
\newcommand\epsop[2]{\epsilon^+({#1},{#2})}
\newcommand\epsopm[2]{\epsilon^\pm({#1},{#2})}

\newcommand\erf[1] {Eq.\ (\ref{#1})}

\def\fB            {\mathfrak{B}}

\def\Gh            {\hat{G}}

\def\gfG           {G}
\def\gfH           {H}

\def\h             {{1/2}}

\def\Hh            {\hat{H}}
\def\Hom           {\mathrm{Hom}}

\newcommand\ind[2] {\mathrm{ind}^{#1}_{#2}}

\def\id            {\mathrm{id}}
\def\Io            {I_\circ}
\def\Jz            {\mathcal{J}_z}

\def\la            {\langle}

\newcommand\labl[1]{\label{#1}\ee}
\newcommand\lablth[1]{\label{#1}}
\newcommand\lablsec[1]{\label{#1}}
\def\lambdab       {\overline{\lambda}}

\newcommand\ls[1]  {[\lambda_{{#1}}]}

\def\LSUn          {\mathit{LSU}(n)}

\def\LTSMO         {[\Delta]_\cM^{(0)}(\Io)}
\def\LTSN          {[\Delta]_\cN(\Io)}

\def\MI            {{M(I)}}
\def\MIo           {{M(\Io)}}

\def\mub           {\overline{\mu}}

\newcommand\N[3]   {N_{{#1},{#2}}^{{#3}}}
\def\NI            {{N(I)}}
\def\NIo           {{N(\Io)}}

\def\nub           {\overline{\nu}}

\def\pio           {\pi_0}
\def\pioi          {\pi_0^{-1}}

\def\QbI           {Q_{\beta;{\Io},{I_1}}}
\def\Qbm           {Q_{\beta,-}}

\def\ra            {\rangle}
\newcommand\res[2] {\mathrm{res}^{#1}_{#2}}

\def\Sect          {\mathrm{Sect}}

\def\sib           {{\sigma_\beta}}

\def\SUd           {\mathit{SU}(3)}
\def\SUn           {\mathit{SU}(n)}
\def\SUz           {\mathit{SU}(2)}

\def\To            {T_\circ}

\def\Ulm           {U_{\lambda,-}}

\def\Ulpm          {U_{\lambda,\pm}}

\def\Ump           {U_{\mu,+}}
\def\Umpm          {U_{\mu,\pm}}

\def\Unpm          {U_{\nu,\pm}}

\def\Urpm          {U_{\rho,\pm}}

\def\ucm           {u_{\canr,-}}
\def\ucp           {u_{\canr,+}}
\def\ucI           {u_{\canr;\Io,I_1}}

\def\umpm          {u_{\mu,\pm}}

\def\vac           {\Omega}

\def\per           {positive energy representation}

\newcommand\ddE[1] {$\mathrm{E}_{#1}$}

\def\Deven         {$\mathrm{D}_{\mathrm{even}}$}

\def\dree          {$\mathrm{III}_1$}

\documentclass[11pt]{article}
\usepackage{amssymb,amsfonts}
\begin{document}


\title{Modular Invariants, Graphs and $\alpha$-Induction for
Nets of Subfactors I}
\author{{\sc Jens B\"ockenhauer} and {\sc David E. Evans}\\ \\
University of Wales Swansea\\
Department of Mathematics\\Singleton Park\\Swansea SA2 8PP}
\maketitle
\nopagebreak

\begin{abstract}
We analyze the induction and restriction of sectors for nets of
subfactors defined by Longo and Rehren. Picking a local subfactor
we derive a formula which specifies the structure of the induced
sectors in terms of the original DHR sectors of the smaller net
and canonical endomorphisms. We also obtain a reciprocity formula
for induction and restriction of sectors, and we prove a certain
homomorphism property of the induction mapping.

Developing further some ideas of F.\ Xu we will apply this theory
in a forthcoming paper to nets of subfactors arising from conformal
field theory, in particular those coming from conformal embeddings
or orbifold inclusions of $\SUn$ WZW models. This will provide a
better understanding of the labeling of modular invariants by
certain graphs, in particular of the A-D-E classification of $\SUz$
modular invariants.
\end{abstract}

\tableofcontents


\newtheorem{definition}{Definition}[section]
\newtheorem{lemma}[definition]{Lemma}
\newtheorem{corollary}[definition]{Corollary}
\newtheorem{theorem}[definition]{Theorem}
\newtheorem{proposition}[definition]{Proposition}


\section{Introduction}

Modular invariants associated to $\SUz$ characters have been
classified by \cite{caiz}, each being labeled by a graph,
a Dynkin Diagram of type A-D-E. Similarly subfactors give
rise to natural invariants, e.g.\ their principal graphs.
Each A, \Deven, $\mathrm{E}_\mathrm{even}$ is the principal
graph (or fusion graph) of a subfactor of index less than four.
Here we begin to look systematically at this relation between
modular invariants, graphs and subfactors.
Our treatment begins with the formulae for the
extension ($\lambda\mapsto\ala$) and the restriction
endomorphism ($\beta\mapsto\sib$) for nets of subfactors
$\cN\subset\cM$ defined by Longo and Rehren \cite{lore}.
We derive several properties of these
extension and restriction endomorphisms, including a
reciprocity formula, and therefore we prefer the names
$\alpha$-induced and $\sigma$-restricted endomorphisms.

We apply the procedure of $\alpha$-induction
to several nets of subfactors arising from conformal field
theory. We pay special attention to the current
algebras of the $\SUn_k$ WZW models. There we are dealing
with nets of subfactors $\cN\subset\cM$ where the smaller
net $\cN$ is given in terms of representations of local loop
groups of $\SUn$. Firstly, we consider conformal embeddings
of type $\SUn_k\subset G_1$ with $G$ simple. In this case the
enveloping net $\cM$ is given by the local loop groups of $G$
in the level $1$ vacuum representation. To such a conformal
embedding corresponds a modular invariant. Secondly, we consider
modular invariants of orbifold type. In this case we can
construct the enveloping net $\cM$ as an extension of $\cN$ by
simple currents; this crossed product construction is similar
to the construction of the field algebra in \cite{dhr1b}.
Our treatment gives some new insights in the programme of
labeling (block-diagonal) modular invariants by certain graphs
initiated by Di Francesco and Zuber \cite{frzu1,frzu2}
(see also \cite{franc}). With $\lambda$ being the localized
endomorphisms associated to the \per s of $\LSUn$ at level $k$
we obtain a fusion algebra generated by the subsectors of the
$\alpha$-induced endomorphisms $\ala$. Graphs are obtained by
drawing the fusion graphs of the $\alpha$-induced endomorphisms
associated to the fundamental representation(s). They
satisfy the axioms for graphs which Di Francesco and
Zuber associate to modular invariants \cite{frzu1}
(see also \cite{pezu2}), and for all our ($\SUz$ and
$\SUd$) examples we reproduce in fact their graphs. For $\SUz$ our
theory yields in fact an explanation why the entries in the
(non-trivial) block-diagonal modular invariants correspond to Coxeter
exponents of the \Deven, \ddE 6 and \ddE 8 Dynkin diagrams.
We will also discuss the application of $\alpha$-induction to extended
$U(1)$ theories from \cite{bumt} and to the minimal models.

In \cite{xu1}, Xu defined a map $\lambda\mapsto a_\lambda$ by a similar,
but different formula for the induced endomorphism. (In fact in his
setting both $\lambda$ and $a_\lambda$ are endomorphisms of the same
\dree -factor $M$.) He has already obtained the fusion graphs for the
conformal inclusions involving $\SUn$, however, we can also treat the
orbifold inclusions of $\SUn$. Our underlying framework is
more general because it applies, for a given net of subfactors
$\cN\subset\cM$ satisfying certain assumptions (which are
fulfilled for many chiral conformal field theory models),
to the whole class of localized, transportable
endomorphisms of $\cN$ whereas Xu restricts his analysis to
the $\LSUn$ setting. Moreover, we believe that our formalism is
more appropriate as the nature of induction and restriction of
sectors becomes more transparent, and we believe that our setting
enables us to present simpler proofs.

This article is the first in a series of papers about modular
invariants, graphs, and nets of subfactors. Here we develop
the machinery of $\alpha$-in\-duc\-tion in a general setting.
In Section \ref{prelim} we derive the braiding fusion equations
that arise naturally from the notion of localized transportable
endomorphisms of algebraic quantum field theory, and which play
a crucial role in our analysis. In Section \ref{aind} we give
the definition and prove several properties of $\alpha$-induction;
we derive an important formula and the homomorphism property
of $\alpha$-induction, and we also establish
$\alpha\sigma$-reciprocity of $\alpha$-induction
and $\sigma$-restriction. 
The game of $\alpha$-induction and $\sigma$-restriction of sectors
generalizes the restriction and (Mackey) induction of group
representations to nets of subfactors which are in general not
governed by group symmetries. Nevertheless, as an illustration we
briefly discuss the case of a net of subfactors arising from a
subgroup of a finite group in Subsection \ref{subgrp}.
In a forthcoming paper \cite{boev2} we will present the above mentioned
applications of this theory to several models of conformal field theory.

\section{Preliminaries}
\lablsec{prelim}

In this section we review several facts about subfactors, sectors,
algebraic quantum field theory and nets of subfactors, which we
will need for our analysis.

\subsection{Subfactors and sectors}

We first briefly review some basic facts about subfactors and Longo's
theory of sectors. For a detailed treatment of these topics we refer
to textbooks on operator algebras, e.g.\ \cite{evka}.

A von Neumann algebra is a weakly closed subalgebra $M\subset\fB(\cH)$
of the algebra of bounded operators on some Hilbert space $\cH$. It is
called a {\em factor} if its center is trivial, $M'\cap M=\bbC \bfe$.
A factor is called {\em infinite} if there is an isometry $v\in M$ with
range projection $vv^*\neq\bfe$, and {\em purely infinite} or
type III if $M_p=pMp$ is infinite for every non-zero projection
$p\in M$.

An inclusion $N\subset M$ of factors with common unit is called a
{\em subfactor}. A subfactor is called irreducible if the relative
commutant is trivial, $N'\cap M=\bbC \bfe$, and it is called
infinite if $N$ and $M$ are infinite factors.
Let $N\subset M$ be an infinite subfactor
on a separable Hilbert space $\cH$. Then there is a vector
$\Phi\in\cH$ which is cyclic and separating for both $M$ and $N$.
Let $J_M$ and $J_N$ be the modular conjugations of $M$ and $N$ with
respect to $\Phi$. Then the endomorphism
\[ \can = \Ad (J_NJ_M)|_M \]
of $M$ satisfies $\can(M)\subset N$ and is called a
{\em canonical endomorphism} from $M$ into $N$. It is unique up to
conjugation by a unitary in $N$. The restriction $\canr=\can|_N$ is
called a dual canonical endomorphism. If the Kosaki index \cite{kosa}
is finite, $[M:N]<\infty$, then there are isometries $v\in M$ and
$w\in N$ such that
\[\bearrl
vm & = \can(m)v\,, \qquad m\in M\,,\\[.5em]
wn & = \canr(n)w\,, \qquad n\in N\,,\\[.5em]
w^*v & = [M:N]^{-\h} \,\bfe = w^* \can(v)
\eear\]
Then $E^M_N(m)=w^*\can(m)w$, $m\in M$, is a conditional expectation
from $M$ onto $N$ and the identity
\[ m=[M:N] \cdot E^M_N(mv^*)v \,, \qquad m\in M\,,\]
holds \cite{lon2}. This means in particular that every $m\in M$
can be written as $m=nv$ for some $n\in N$, i.e.\ $M=Nv$.

For any unital ${}^*$-algebra $M$ we denote by $\End(M)$
the set of unital ${}^*$-en\-do\-morphisms of $M$.
For $\lambda,\mu\in\End(M)$ we define the intertwiner space
\[ \Hom_M(\lambda,\mu)=
\{ t\in M : \,\, t\lambda(m)=\mu(m)t \,,\quad m\in M \} \]
and
\[ \la \lambda,\mu \ra_M = \dim \, \Hom_M(\lambda,\mu) \,. \]
We have $\la\lambda,\mu\ra_M=\la \mu,\lambda \ra_M$.
Now let $M$ be a type III factor. 
An endomorphism $\lambda\in\End(M)$ is called {\em irreducible} if
$\lambda(M)'\cap M=\bbC\bfe$.
Endomorphisms $\lambda,\mu\in\End(M)$ are called
(inner) {\em equivalent} if there is a unitary $u\in M$ such that
$\lambda=\Ad(u)\circ\mu$. The quotient of $\End(M)$ by inner
equivalence is called the set of {\em sectors} of $M$ and denoted
by $\Sect(M)$, and the equivalence class of $\lambda\in\End(M)$
is denoted by $[\lambda]_M$. However, we often drop the suffix
and write $[\lambda]$ for $[\lambda]_M$ as long as it is clear
which factor is meant.
There is a natural product of sectors coming from
the composition of endomorphisms. Explicitly,
$[\lambda]\times[\mu]=[\lambda\circ\mu]$. There is also an addition
of sectors. Let $\lambda_i\in\End(M)$, $i=1,2,...,n$.
Since $M$ is infinite we can take a set of isometries
$t_i\in M$, $i=1,2,...,n$,
satisfying the relations of the Cuntz algebra $\cO_n$,
\[ t_i^*t_j = \del ij \, \bfe \,,\qquad
\sum_{i=1}^n t_it_i^* = \bfe \,. \]
Define $\lambda\in\End(M)$ by
\[ \lambda(m) = \sum_{i=1}^n t_i \, \lambda_i(m) \, t_i^*
\,, \qquad m\in M\,. \]
Then $[\lambda]$ does not depend on the choice of the set of
isometries and hence we can define the sum
\[ \bigoplus_{i=1}^n \,\, [\lambda_i] = [\lambda] \,. \]
Each $[\lambda_i]$ is called a {\em subsector} of $[\lambda]$.
With the operations $\times$ and $\oplus$ that fulfill
associativity and distributivity, $\Sect(M)$ becomes a unital
semi-ring, and the unit is given by the identity
(or trivial) sector $[\id]$.

For $\lambda\in\End(M)$ irreducible $\lambdab\in\End(M)$ is called
conjugate if $[\lambda\circ\lambdab]$ and $[\lambdab\circ\lambda]$
both contain the identity sector once. The conjugate is unique up
to inner equivalence. For general $\lambda$ let $\can_\lambda$
be the canonical endomorphism of $M$ into $\lambda(M)$. Then a
conjugate is given by $\lambdab=\lambda^{-1}\circ\can_\lambda$.
$[\lambdab]$ is called the conjugate sector, and the map
$[\lambda]\mapsto[\lambdab]$ preserves sums
(if $[\lambda]=[\lambda_1]\oplus[\lambda_2]$ then
$[\lambdab]=[\lambdab_1]\oplus[\lambdab_2]$) and
reverses products (if $[\lambda]=[\mu]\times[\nu]$ then
$[\lambdab]=[\nub]\times[\mub]$). Furthermore,
for an automorphism $\alpha\in\Aut(M)$ we have
$[\alpha^{-1}]=[\overline{\alpha}]$. The number
$d_\lambda=[M:\lambda(M)]^\h$ is called the
{\em statistical dimension} of $\lambda$. Then
$d_\lambda=d_{\lambda'}$ if $[\lambda]=[\lambda']$, and
$d_\lambda=d_{\lambdab}$. For
$\lambda_1,\lambda_2\in\End(M)$ such that
$[\lambda]=[\lambda_1]\oplus[\lambda_2]$ we have
\[ \la\lambda,\mu\ra_M=\la\lambda_1,\mu\ra_M+\la\lambda_2,\mu\ra_M \,.\]
If $\lambda,\mu,\nu,\lambdab,\mub\in\End(M)$ have finite
statistical dimension and $\lambdab$ and $\mub$
are conjugates of $\lambda$ and $\mu$, respectively, then we have
\cite{lon4}
\be
\la \lambda\circ\mu,\nu \ra_M = \la \lambda,\nu\circ\mub \ra_M
= \la \mu,\lambdab\circ\nu \ra_M \,,
\labl{frod}
in particular $\la\lambda,\mu\ra_M=\la\lambdab,\mub\ra_M$.

\subsection{Statistics operators in algebraic quantum field theory}

Let us briefly review some facts about the algebraic framework of
quantum field theory \cite{dhr1,dhr1b,dhr2,dhr2b,haag}.
As all our later applications are chiral theories we present
the whole setting with the unit circle $S^1$ as the underlying
``space-time'' from the beginning. Since we will make explicit use
of several well-known results and in order to make this article more
self-contained we prefer to present the proofs which are simple
and instructive, but compare also \cite{frs1,frs2}.
Fix a point $z\in S^1$ on the circle and set
\[ \Jz = \{ I\subset S^1 \,\,\,\mbox{non-void open interval}
\,,\,\,\,z\notin \bar{I} \} \,, \]
where $\bar{I}$ denotes the closure of $I$.
A Haag-Kastler net on the punctured circle
$\cA=\{A(I)\,,\,\,\,I\in\Jz\}$ is a family of von
Neumann algebras acting on a Hilbert space $\cH_0$ such that
isotony holds, i.e.\ $I\subset J$ implies $A(I)\subset A(J)$,
and we also have locality, i.e.\ $I_1\cap I_2=\emptyset$
implies $A(I_1)\subset A(I_2)'$. For subsets $R\subset S^1$
(which may touch or contain the ``point at infinity'' $z$)
we define
\[ \CAo (R)=\bigcup_{J\in\Jz\,,\,\,\,J\subset R} A(J) \,,\qquad
\CA(R)= \overline{\CAo(R)}^{\|\cdot\|} \,. \]
As usual, we denote the $C^*$-algebra of the whole circle by the same
symbol as the net itself, $\cA=\CA(S^1)$. An endomorphism
$\lambda\in\End(\cA)$ is called {\em localized} in an interval
$I\in\Jz$ if $\lambda(a)=a$ for all $a\in\CA(I')$, where $I'$
denotes the interior of the complement of $I$. A localized
endomorphism $\lambda$ is called {\em transportable} if
for all $J\in\Jz$ there are unitaries $U_{\lambda;I,J}\in\cA$,
called {\em charge transporters}, such that
$\tilde{\lambda}=\Ad(U_{\lambda;I,J})\circ\lambda$ is
localized in $J$. By $\Delta_\cA(I)$
we denote the set of localized transportable (``DHR'')
endomorphisms of $\cA$ localized in $I\in\Jz$.

Let us now assume {\em Haag duality} (on the punctured circle),
\be
\cA(I)=\CA(I')' \,,\qquad I\in\Jz \,.
\labl{Haag}
Note that then an
endomorphism $\lambda\in\DelAIo$ leaves any local algebra
$A(K)$ with $K\in\Jz$, $I\subset K$, invariant since
$a'\lambda(a)=\lambda(a'a)=\lambda(aa')=\lambda(a)a'$
for any $a\in A(K)$ and $a'\in \CA(K')$,
hence $\lambda(a)\in A(K)$ by Haag duality.

\begin{lemma}
Let $I_1,I_2\in\Jz$ such that $I_1\cap I_2=\emptyset$ and
let $\lambda_i\in\Delta_\cA(I_i)$, $i=1,2$. Then
$\lambda_1$ and $\lambda_2$ commute,
$\lambda_1\circ\lambda_2=\lambda_2\circ\lambda_1$.
\end{lemma}

\bproof
Take $I\in\Jz$ arbitrary. Then choose intervals $J_1,J_2\in\Jz$
such that $J_i\cap I=\emptyset$, $i=1,2$, and that there are also
intervals $K_1,K_2\in\Jz$, $K_i\supset I_i\cup J_i$, $i=1,2$, and
$K_1\cap K_2=\emptyset$. By transportability there are unitaries
$U_i\equiv U_{\lambda_i;I_i,J_i}$ such that
$\tilde{\lambda}_i=\Ad (U_i)\circ\lambda_i \in\Delta_\cA(J_i)$,
$i=1,2$. Then $U_i\in A(K_i)$ by Haag duality, hence
$U_1U_2=U_2U_1$ and $\tilde{\lambda}_1(U_2)=U_2$ and
$\tilde{\lambda}_2(U_1)=U_1$. Then for any $a\in A(I)$ we
have $\tilde{\lambda}_i(a)=a$, $i=1,2$, and thus
\[ \bearll
\lambda_1\circ\lambda_2 (a)
&= \Ad (U_1^*)\circ\tilde{\lambda}_1\circ \Ad (U_2^*)
\circ\tilde{\lambda}_2 (a) \\[.4em]
&= \Ad (U_1^*\tilde{\lambda}_1(U_2^*)) \circ \tilde{\lambda}_1
\circ \tilde{\lambda}_2(a) \\[.4em]
&= U_1^* U_2^* a U_2 U_1 =  U_2^* U_1^* a U_1 U_2 \\[.4em]
&= \Ad (U_2^*\tilde{\lambda}_2(U_1^*)) \circ \tilde{\lambda}_2
\circ \tilde{\lambda}_1(a) \\[.4em]
&= \Ad (U_2^*)\circ\tilde{\lambda}_2\circ \Ad (U_1^*)
\circ\tilde{\lambda}_1 (a) \\[.4em]
&= \lambda_2\circ\lambda_1 (a) \,.
\eear \]
Since $I$ was arbitrary it follows
$\lambda_1\circ\lambda_2(a)=\lambda_2\circ\lambda_1(a)$
for any $a\in\cA$.
\eproof

Now assume that $\lambda,\mu$ are localized in the same interval
$I\in\Jz$, $\lambda,\mu\in\Delta_\cA(I)$. Then they will in general not
commute, however, they are intertwined by a unitary operator which
will be discussed in the following. Choose $I_1,I_2\in\Jz$ such that
$I_1\cap I_2=\emptyset$. Then there are unitaries
$U_1\equiv U_{\lambda;I,I_1}$ and $U_2\equiv U_{\mu;I,I_2}$
such that $\lambda_1=\Ad(U_1)\circ\lambda\in\Delta_\cA(I_1)$
and $\mu_2=\Ad(U_2)\circ\mu\in\Delta_\cA(I_2)$. We set
\[ \epsilon_{U_1,U_2}^{I_1,I_2} (\lambda,\mu) =
\mu(U_1^*)U_2^*U_1\lambda(U_2) \,. \]
This operator has remarkable invariance properties. Let
\[ \cJ_{z,\mathrm{dis}}^2 = \{ (I_1,I_2) \in \Jz \times
\Jz \,,\,\,\, I_1\cap I_2 = \emptyset \} \,. \]
For disjoint intervals $I_1,I_2\in\Jz$ denote $I_2>I_1$
(respectively $I_2<I_1$) if $I_1$ lies clockwise
(respectively counter-clockwise) to $I_2$ relative
to the point $z$. Let
\[ \bearll
\cJ_{z,+}^2 &=  \{ (I_1,I_2)\in \Jz \times
\Jz \,,\,\,\, I_2>I_1 \} \,,\\[.4em]
\cJ_{z,-}^2 &= \{ (I_1,I_2)\in \Jz \times
\Jz \,,\,\,\, I_2<I_1 \} \,.
\eear \]
Then clearly $\cJ_{z,\mathrm{dis}}^2 = \cJ_{z,+}^2 \cup \cJ_{z,-}^2$.

\begin{lemma}
The operators $\epsilon_{U_1,U_2}^{I_1,I_2} (\lambda,\mu)$
do not depend on the special choice of $U_1$ and $U_2$,
moreover, varying $I_1$ and $I_2$,
$\epsilon_{U_1,U_2}^{I_1,I_2} (\lambda,\mu)$
remains constant on $\cJ_{z,+}^2$ and $\cJ_{z,-}^2$.
\lablth{einv}
\end{lemma}

\bproof
First replace $U_1$ by $\tilde{U}_1$ such that
$\tilde{\lambda}_1=\Ad(\tilde{U}_1)\circ\lambda\in\Delta_\cA(I_1)$
as well. Then with $V_1=\tilde{U}_1 U_1^*$ we have
$\tilde{\lambda}_1=\Ad (V_1)\circ \lambda_1$ and hence
$V_1\in A(I_1)$ by Haag duality. Then we have
\[ \bearll
\epsilon_{U_1,\tilde{U}_2}^{I_1,I_2} (\lambda,\mu)
&= \mu (\tilde{U}_1^*)U_2^*\tilde{U}_1\lambda(U_2) \\[.4em]
&= \mu (U_1^*V_1^*)U_2^*V_1U_1\lambda(U_2) \\[.4em]
&= \mu (U_1^*) U_2^* \mu_2(V_1^*) V_1 U_1 \lambda(U_2) \\[.4em]
&= \epsilon_{U_1,U_2}^{I_1,I_2} (\lambda,\mu) \,,
\eear \]
since $\mu_2(V_1)=V_1$ by $I_1\cap I_2=\emptyset$. In the same
way we can replace $U_2$ by some $\tilde{U}_2$ such that
$\tilde{\mu}_2=\Ad(\tilde{U}_2)\circ\mu\in\Delta_\cA(I_2)$.
In the next step we replace $I_1$ by some $\tilde{I_1}$ such
that $\tilde{I}_1\cap I_1\neq\emptyset$ but still
$\tilde{I}_1\cap I_2=\emptyset$. We can now assume that our
chosen $\tilde{U}_1$ is such that
$\tilde{\lambda}_1\in\Delta_\cA(\tilde{I}_1\cap I_1)$, and hence
we can use the same $\tilde{U}_1$ for the new interval $\tilde{I}_1$.
In the same way we can replace $I_2$ by $\tilde{I}_2$. As long
as $\tilde{I}_1\cap\tilde{I}_2=\emptyset$ we have the freedom to
vary $\tilde{U}_1$ and $\tilde{U}_2$, and so on. Now assume
that we have $I_2>I_1$ for our initial intervals. By iteration of
the above arguments we can reach any pair of intervals
in $\cJ_{z,+}^2$, and similarly in $\cJ_{z,-}^2$ if $I_1<I_2$,
the lemma is proven.
\eproof

We conclude that for any $\lambda,\mu\in\DelAI$ there are only
two operators
$\epsopm \lambda\mu=\epsilon_{U_1,U_2}^{I_1,I_2} (\lambda,\mu)$,
where $(I_1,I_2)\in\cJ_{z,\pm}^2$, but $\epsop \lambda\mu$ and
$\epsom \lambda\mu$ may be different in general. We now have
even the choice to set $I_1=I$ and $U_1=\bfe$. We choose intervals
$I_\pm\in\Jz$ such that $I_+>I$ and $I_-<I$. If
$\Umpm\equiv U_{\mu;I,I_\pm}$ are unitaries such that
$\mu_\pm=\Ad (\Umpm)\circ\mu\in\Delta_\cA(I_\pm)$ then we find
by putting $I_2=I_+$ or $I_2=I_-$
\[ \epsopm \lambda\mu = \Umpm^* \, \lambda (\Umpm) \,. \]
The $\epsopm \lambda\mu$'s are usually called
{\em statistics operators}.
Choose $K_+,K_-\in\Jz$ such that $I\cup I_\pm\subset K_\pm$
and $I_\pm\cap K_\mp=\emptyset$. Note that
$\Umpm\in A(K_\pm)$ by Haag duality.

\begin{lemma}
For $\lambda,\mu,\nu\in\DelAI$ we have
\bea
\epsopm \lambda\mu \cdot \lambda\circ\mu (a) &=& \mu\circ\lambda (a)
\cdot \epsopm \lambda\mu \,, \qquad a\in\cA\,,
\label{zwei} \\
\epsopm \lambda\mu &\in& A(I)\,,
\label{drei} \\
\epsop \lambda\mu &=& (\epsom \mu\lambda )^* \,,
\label{vier} \\
\epsopm {\lambda\circ\mu}\nu &=& \epsopm \lambda\nu \,
\lambda (\epsopm \mu\nu) \,,
\label{fuenf} \\
\epsopm \lambda{\mu\circ\nu} &=& \mu (\epsopm \lambda\nu) \,
\epsopm \lambda\mu \,.
\label{sechs}
\eea
\lablth{stat}
\end{lemma}

\bproof
Ad \erf{zwei}: For $a\in\cA$ we compute
\[ \bearll
\epsopm \lambda\mu \cdot \lambda\circ\mu(a) &=
\Umpm^* \, \lambda (\Umpm) \cdot \lambda\circ\mu(a) \\[.4em]
&= \Umpm^* \, \lambda(\Umpm \, \mu(a)) \\[.4em]
&= \Umpm^* \, \lambda ( \mu_\pm(a)\, \Umpm) \\[.4em]
&= \Umpm^* \cdot \mu_\pm \circ \lambda(a) \cdot \lambda (\Umpm) \\[.4em]
&= \mu\circ\lambda(a) \cdot \Umpm^* \, \lambda (\Umpm) \\[.4em]
&= \mu\circ\lambda(a) \cdot \epsopm \lambda\mu \,. \eear \]

Ad \erf{drei}: For $a\in\cC_\cA(I')$ \erf{zwei} reads
$\epsopm \lambda\mu \, a = a \, \epsopm \lambda\mu$, i.e.\
$\epsopm \lambda\mu \in \cC_\cA(I')'=A(I)$.

Ad \erf{vier}: From $\Umpm\in A(K_\pm)$ it follows
$\lambda_-(\Ump)=\Ump$ and $\mu_+(\Ulm)=\Ulm$. Hence
\[ \bearll
\epsop \lambda\mu &=
\Ump^* \, \lambda (\Ump)  \\[.4em]
&= \Ump^* \, \Ulm^* \, \Ulm \, \lambda(\Ump) \\[.4em]
&= \Ump^* \, \Ulm^* \, \lambda_-(\Ump) \, \Ulm \\[.4em]
&= \Ump^* \, \Ulm^* \, \Ump \, \Ulm \\[.4em]
&= \Ump^* \, \mu_+(\Ulm^*) \, \Ump \, \Ulm \\[.4em]
&= \Ump^* \, \Ump \, \mu (\Ulm^*) \, \Ulm \\[.4em]
&= \mu (\Ulm^*) \, \Ulm \\[.4em]
&= (\epsom \mu\lambda)^* \,. \eear \]

Ad \erf{fuenf}: Clearly $(\lambda\circ\mu)_\pm\in\Delta_\cA(I_\pm)$
where
\[ (\lambda\circ\mu)_\pm = \lambda_\pm\circ\mu_\pm =
\Ad (U_{\lambda\circ\mu,\pm}) \circ\lambda\circ\mu\,,\qquad
U_{\lambda\circ\mu,\pm} = \Ulpm \, \lambda (\Umpm) \,. \]
Hence
\[ \bearll
\epsopm {\lambda\circ\mu}\nu &=
\Unpm^* \cdot \lambda\circ\mu (\Unpm)  \\[.4em]
&=  \Unpm^* \, \lambda(\Unpm) \, \lambda(\Unpm^*)
\cdot \lambda\circ\mu (\Unpm) \\[.4em]
&= \epsopm \lambda\nu \, \lambda (\epsopm \mu\nu) \,. \eear \]

Ad \erf{sechs}: This follows now easily from Eqs.\ (\ref{vier})
and (\ref{fuenf}).
\eproof

Note that \erf{vier} nicely reflects the invariance properties
of $\epsopm \lambda\mu$ as stated in Lemma \ref{einv}.

\subsection{The braiding fusion equations}
\lablsec{bfe}

We will now describe how the naturality and braiding fusion equations
(BFEs) arise in the algebraic framework. The content of this
subsection is not essentially new (e.g.\ versions of these equations
have already been given in \cite{frs2}), however, as we will make
explicit use of the different versions of the BFE we again
present the proofs. Moreover, in view of our applications we want
to formulate the BFEs for local intertwiners and therefore we
have to require strong additivity of the underlying Haag-Kastler
net. Strong additivity (or ``irrelevance of points'') means that
$A(I)=A(I_1)\vee A(I_2)$ whenever intervals $I_1$ and $I_2$ are
obtained by removing one single point from the interval $I\in\Jz$.
This requirement basically ensures the equivalence of local
and global intertwiners. 
In the following we will often consider
elements of the set $\DelAI$ as elements of $\End(A(K))$ 
for $I,K\in\Jz$ such that $I\subset K$ which is possible
since elements of $\DelAI$ leave $A(K)$ invariant.

\begin{lemma}
Suppose that $\cA$ is strongly additive. Then for
$\lambda,\mu\in\DelAIo$, $\Io\in\Jz$, we have
\be
\Hom_\cA(\lambda,\mu)=\Hom_\AIo(\lambda,\mu) \,.
\ee
\lablth{globloc}
\end{lemma}

\bproof
We first show ``$\subset$''. Assume $T\in\Hom_\cA(\lambda,\mu)$.
Then clearly $T\lambda(a)=\mu(a)T$ for all $a\in\AIo$.
Moreover, as $Ta=T\lambda(a)=\mu(a)T=aT$
for all $a\in \CA(\Io')$ we find $T\in\CA(\Io')'=A(\Io)$,
proving ``$\subset$''. 

Next we show ``$\supset$''. Assume
$\To\in\Hom_\AIo(\lambda,\mu)$. It suffices to show
$\To\lambda(a)=\mu(a)\To$ for all $a\in\AI$ and all
$I\in\Jz$ such that $\Io\subset I$ ($\Io\neq I$) because
then $\To\in\Hom_\cA(\lambda,\mu)$ by norm continuity.
First assume that $\Io$ and $I$ have
one boundary point in common, i.e.\ $I$ extends $\Io$ one
one side. Then $I_1=I\cap\Io'$ is an interval in $\Jz$ and
$\AI=\AIo\vee A(I_1)$ by strong additivity.  We have
$\To\lambda(a)=\mu(a)\To$ for all $a\in\AIo$ by assumption
and also $\To\lambda(a)=\To a=a\To=\mu(a)\To$ for all
$a\in A(I_1)$ since $\To\in\AIo$. Hence $\To$ intertwines
$\lambda$ and $\mu$ on the subalgebra of $\AI$ which is
algebraically generated by $\AIo$ and $A(I_1)$ and is weakly
dense by strong additivity. As endomorphisms in $\DelAIo$
are weakly continuous on any $A(I)$, $\Io\subset I$, it
follows $\To\lambda(a)=\mu(a)\To$ for all $a\in\AI$.
If $I$ has no common boundary point with $\Io$ we just
have to repeat the procedure to extend the interval also
on the other side.
\eproof

Now we are ready to prove the naturality equations for local
intertwiners.

\begin{proposition}
For $\lambda,\mu,\rho\in\DelAIo$, $\Io\in\Jz$,
and $T\in \Hom_\AIo (\lambda,\mu)$
we have the naturality equations
\bea
\rho(T) \, \epsopm \lambda \rho &=&
\epsopm \mu \rho \, T \,, \label{nat2} \\
T \, \epsopm \rho \lambda &=&
\epsopm \rho \mu \, \rho (T) \,. \label{nat1}
\eea
\end{proposition}

\bproof
Choose intervals $I_+,I_-\in\Jz$
such that $I_-<\Io<I_+$.
We take unitaries $\Urpm\in\cA$ such that
$\rho_\pm=\Ad (\Urpm)\circ\rho$ are localized in $I_\pm$.
Then $T\lambda(\Urpm)=\mu(\Urpm)T$ by Lemma \ref{globloc}.
Moreover, $\rho_\pm(T)=T$ as $T\in\AIo$.
We can now compute
\[ \bearll
\rho(T) \, \epsopm \lambda \rho
&= \rho(T) \, \Urpm^* \, \lambda (\Urpm) \\[.4em]
&= \Urpm^* \, \rho_\pm(T) \, \lambda (\Urpm) \\[.4em]
&= \Urpm^* \, T \, \lambda (\Urpm) \\[.4em]
&= \Urpm^* \, \mu (\Urpm) \, T \\[.4em]
&= \epsopm \mu\rho \, T \,, \eear \]
and \erf{nat1} is obtained just by applying \erf{nat2} to
$T^*\in \Hom_\AIo (\mu,\lambda)$ and using \erf{vier}.
\eproof

By use of Eqs.\ (\ref{fuenf}) and (\ref{sechs}) we obtain immediately
the following

\begin{corollary}
For $\lambda,\mu,\nu,\rho\in\DelAIo$, $\Io\in\Jz$, and
$S\in \Hom_\AIo (\lambda\circ\mu,\nu)$ we have the BFEs
\bea
\rho(S) \, \epsopm \lambda \rho \lambda (\epsopm \mu \rho)
&=& \epsopm \nu \rho \, S \,, \label{BFE2} \\
S \, \lambda (\epsopm \rho \mu) \epsopm \rho \lambda
&=& \epsopm \rho \nu \, \rho (S) \,. \label{BFE1}
\eea
\end{corollary}
By Lemma \ref{stat}, Eqs.\ (\ref{zwei}) and (\ref{drei}), we find
$\epsopm\lambda\mu\in\Hom_\AIo(\lambda\circ\mu,\mu\circ\lambda)$.
Using \erf{BFE2} and also \erf{fuenf} we obtain the
Yang-Baxter equation (YBE).

\begin{corollary}
For $\lambda,\mu,\nu\in\DelAIo$ we have the YBE
\be
\nu(\epsopm\lambda\mu) \, \epsopm\lambda\nu \, \lambda(\epsopm\mu\nu) =
\epsopm\mu\nu \, \mu(\epsopm\lambda\nu) \, \epsopm\lambda\mu \,.
\labl{YBEoe}
\lablth{YBE}
\end{corollary}
We remark that the YBE is also true without the assumption of
strong additivity because the statistics operators are global
intertwiners.

Assume we have a Haag-Kastler net $\cN=\{N(I)\,,\,\,\,I\in\Jz\}$
of von Neumann algebras acting on a Hilbert space $\cH$. If (the
$C^*$-algebra) $\cN$ leaves a subspace $\cH_0\subset\cH$ invariant
and the corresponding subrepresentation $\pio$ of the defining
representation of $\cN$ is faithful, we denote by
$\cA=\{A(I)\,,\,\,\,I\in\Jz\}$ the isomorphic net given by
\[ A(I) = \pio(N(I))\,, \qquad I\in\Jz \,.\]
Then strong additivity of the net $\cN$ is equivalent to
strong additivity of the net $\cA$. If the net $\cA$ is
Haag dual the we say that $\cN$ {\em has a faithful Haag dual
subrepresentation}. In that case one checks that
$N(I)=\CN(I')'\cap\cN$ for $I\in\Jz$.

Let $\Delta_\cN(I)$ denote the set of
transportable endomorphisms of $\cN$ localized in $I\in\Jz$, i.e.\
for $\lambda\in\Delta_\cN(I)$ and any $J\in\Jz$ there are
unitary charge transporters
$u_{\lambda;I,J}\in\cN$ such that
$\tilde{\lambda}=\Ad(u_{\lambda;I,J})\circ\lambda$ is
localized in $J$. Then
$U_{\lambda_0;I,J}=\pio(u_{\lambda;I,J})$
is a charge transporter of
\[ \lambda_0 = \pio\circ\lambda\circ\pioi \in \Delta_\cA(I) \,.\]
Note that, if $\cN$ has a Haag dual subrepresentation,
elements of $\Delta_\cN(I)$ leave $N(K)$ invariant whenever
$K\in\Jz$ contains $I$, so that elements of $\Delta_\cN(I)$
can also be considered as elements of $\End(N(K))$.

Now choose again $\Io,I_\pm\in\Jz$ such that $I_-<\Io<I_+$.
For $\lambda,\mu\in\DelNIo$ we set
$\umpm=u_{\mu;\Io,I_\pm}$, and
\[ \epspm \lambda\mu = \umpm^* \lambda (\umpm) \]
so that
\[ \epsopm {\lambda_0}{\mu_0} = \pio (\epspm \lambda\mu) \,. \]
We call the $\epsp \lambda\mu$'s statistics operators as well.

Now assume that $\cN$ is strongly additive and
let $\lambda,\mu,\rho\in\DelNIo$ and
$t\in \Hom_\NIo(\lambda,\mu)$. Then
$T=\pio(t)\in \Hom_\AIo (\lambda_0,\mu_0)$. This way we
obtain $\Hom_\cN(\lambda,\mu)=\Hom_\NIo(\lambda,\mu)$
from Lemma \ref{globloc}, and we have the naturality equations
\[ \bearrl
\rho_0(T) \, \epsopm {\lambda_0}{\rho_0}
&= \,\,\, \epsopm {\mu_0}{\rho_0} \, T \,, \\[.4em]
T \, \epsopm {\rho_0}{\lambda_0}
&= \,\,\, \epsopm {\rho_0}{\mu_0} \, \rho_0 (T) \,.
\eear\]
Applying $\pioi$ to this and Lemma \ref{stat}
we arrive at

\begin{corollary}
Assume that $\cN$ has a faithful Haag dual subrepresentation.
Then we have for $\lambda,\mu\in\DelNIo$, $\Io\in\Jz$,
\bea
\epspm \lambda\mu \cdot \lambda\circ\mu (n) &=& \mu\circ\lambda (n)
\cdot \epspm \lambda\mu \,, \qquad n\in\cN\,,
\label{zwein} \\
\epspm \lambda\mu &\in& N(\Io)\,,
\label{drein} \\
\epsp \lambda\mu &=& (\epsm \mu\lambda )^* \,,
\label{viern} \\
\epspm {\lambda\circ\mu}\nu &=& \epspm \lambda\nu \,
\lambda (\epspm \mu\nu) \,,
\label{fuenfn} \\
\epspm \lambda{\mu\circ\nu} &=& \mu (\epspm \lambda\nu) \,
\epspm \lambda\mu \,.
\label{sechsn}
\eea
If in addition $\cN$ is strongly additive and also
$\nu,\rho\in\DelNIo$, then for $t\in \Hom_\NIo (\lambda,\mu)$
we have the naturality equations
\bea
\rho(t) \, \epspm \lambda \rho &=&
\epspm \mu \rho \, t \,, \label{natn2} \\
t \, \epspm \rho \lambda &=&
\epspm \rho \mu \, \rho (t) \,, \label{natn1}
\eea
for $s\in \Hom_\NIo (\lambda\circ\mu,\nu)$ we have the BFEs
\bea
\rho(s) \, \epspm \lambda \rho \lambda (\epspm \mu \rho)
&=& \epspm \nu \rho \, s \,, \label{BFE4} \\
s \, \lambda (\epspm \rho \mu) \epspm \rho \lambda
&=& \epspm \rho \nu \, \rho (s) \,, \label{BFE3}
\eea
and the YBE
\be
\nu(\epspm\lambda\mu) \, \epspm\lambda\nu \, \lambda(\epspm\mu\nu) =
\epspm\mu\nu \, \mu(\epspm\lambda\nu) \, \epspm\lambda\mu \,.
\labl{YBEe}
\end{corollary}

\subsection{Nets of subfactors}
\lablsec{prelLR}

A net of von Neumann algebras (or even factors)
over a partially ordered index set $\cJ$ is an assignment
$\cM:\cJ\ni i\mapsto M_i$ of von Neumann algebras (or factors) on a
Hilbert space $\cH$ such that we have isotony, $M_i\subset M_j$
whenever $i\le j$. A net of subfactors consists of two nets of factors
$\cN$ and $\cM$ such that we have subfactors $N_i\subset M_i$ for
all $i\in\cJ$. We simply write $\cN\subset\cM$. A net of subfactors
is called standard if there is a vector $\vac\in\cH$ that is cyclic
and separating for every $M_i$ on $\cH$ and $N_i$ on a subspace
$\cH_0\subset\cH$. Note that the projection $e_N\in\fB(\cH)$ onto
$\cH_0$ is the Jones projection for each inclusion $N_i\subset M_i$
for a standard net of subfactors. If there is also an assignment
$E:\cJ\ni i\mapsto E_i$ of faithful normal conditional expectations
from $M_i$ onto $N_i$ such that $E_i=E_j|_{M_i}$ for $i\le j$ then
we say that $\cN\subset\cM$ has a faithful normal
conditional expectation. $E$ is called standard if it
preserves the vector state $\omega=\la\vac,\cdot\,\vac\ra$. If the
index set $\cJ$ is directed we simply say $\cN\subset\cM$ is a
directed net and we can form the $C^*$-algebras
$\overline{\bigcup_{i\in\cJ}N_i}$ and $\overline{\bigcup_{i\in\cJ}M_i}$
and denote it, by abuse of notation, by the same symbols as used
for the nets, $\cN$ and $\cM$, respectively.

In \cite{lore} the following is proven

\begin{proposition}
Let $\cN\subset\cM$ be a directed standard net of subfactors with
a standard conditional expectation. For every $i\in\cJ$ there is
an endomorphism $\can$ of the $C^*$-algebras $\cM$ into $\cN$
such that $\can|_{M_j}$ is a canonical endomorphism of $M_j$ into
$N_j$ whenever $i\le j$. Furthermore, $\can$ acts trivially on
$M_i'\cap\cN$. As $i\in\cJ$ varies to any $i'\in\cJ$ the
corresponding $\can$ and $\can'$ are inner equivalent by a
unitary in $N_k$ provided $i,i'\le k$.
\lablth{LR1}
\end{proposition}

Since (for a fixed $i\in\cJ$) $\can$ is a canonical endomorphism
of $M_j$ into $N_j$ whenever $i\le j$ there is a restriction
of $\can$ to $\cN$ that we denote by $\canr$,
\be
\canr = \can|_\cN \in \End(\cN) \,.
\ee

\begin{proposition}
Let $\cN\subset\cM$ be a directed standard net of subfactors with
a standard conditional expectation. Let $\can\in\End(\cM)$ be
associated with some $i\in\cJ$ and $\canr\in\End(\cN)$ its
restriction as above. Then we have unitary equivalences
\be
\pi^0 \simeq \pio\circ\can \qquad \mbox{and} \qquad
\pi^0 |_\cN \simeq \pio\circ\canr
\ee
where $\pi^0$ is the defining representation of $\cM$ on $\cH$
and $\pio$ the ensuing representation of $\cN$ on
$\cH_0=\overline{\cN\vac}$.
\lablth{LR2}
\end{proposition}

It is also proven in \cite{lore} that the Kosaki index is constant in
a directed standard net of subfactors with a standard conditional
expectation. Moreover, for such nets the following is shown in
\cite{lore}. Pick $\can$ and $\canr$ for some $i\in\cJ$ as above.
Then there is an isometry $w\in N_i$ satisfying $wn=\canr(n)w$ for
all $n\in\cN$ and inducing the conditional expectation $E$ by
$E(m)=w^*\can(m)w$ for $m\in\cM$. If in addition the index is
finite, $[M:N]\equiv[M_i:N_i]<\infty$, then there is also an
isometry $v\in M_i$ satisfying $vm=\can(m)v$ for all $m\in\cM$
and $w^*v=[M:N]^{-1/2}\bfe=w^*\can(v)$. Then clearly
$E(vv^*)=[M:N]^{-1}\bfe$, and we have also $M_j=N_jv$
whenever $i\le j$, and finally $\cM=\cN v$.

A directed standard net of subfactors with a standard conditional
expectation is called a quantum field theoretical net
of subfactors if the index set $\cJ$ admits a causal structure
and we have $N_i\subset M_j'$ if $i$ and $j$ are causally disjoint.
For our purposes we choose the directed set $\cJ=\Jz$ and assume
that we have a given quantum field theoretical net of subfactors
$\cN\subset\cM$. We denote by $\cA$ the net (and the $C^*$-algebra)
\be
A(I) = \pio (N(I)) \,, \qquad I\in\Jz \,.
\ee
As we are dealing with factors, $\pio$ is automatically
faithful. We assume that $\cA$ satisfies Haag duality,
i.e.\ $\cN$ has a faithful Haag dual subrepresentation.

Fix an interval $\Io\in\Jz$ and take the endomorphism
$\can$ of Prop.\ \ref{LR1}. First note
that Proposition \ref{LR2} tells us that $\canr\in\DelNIo$.
Let us consider the situation that $\pi^0$ decomposes into a finite
number of representations of $\cN$ as follows,
\[ \pio\circ\canr \simeq \pi^0|_\cN \simeq \bigoplus_{\ell=0}^n
\,\, m_\ell \, \pi_\ell \]
where $\pi_\ell$, $\ell=0,1,...,n$, are irreducible, mutually disjoint
representations of $\cN$ and $m_\ell$ are multiplicities. Assume that
$\pi_\ell$ are such that we can write
\[ \pio\circ\canr \simeq \bigoplus_{\ell=0}^n \,\, m_\ell \cdot
\pio\circ\lambda_\ell \]
with $\lambda_\ell\in\DelNIo$. Then this means that we have isometries
$T_{\ell,r}\in\fB(\cH_0)$, $\ell=0,1,...,n$, $r=1,2,...,m_\ell$, such
that
\[ T_{\ell,r}^*T_{\ell',r'} = \del \ell{\ell'} \del r{r'}
\bfe \,, \qquad \sum_{\ell=0}^n\sum_{r=1}^{m_\ell}
T_{\ell,r}T_{\ell',r'}^* = \bfe \,, \]
and
\[ \pio\circ\canr(n) = \sum_{\ell=0}^n\sum_{r=1}^{m_\ell}
T_{\ell,r}\cdot \pio\circ\lambda_\ell(n)\cdot T_{\ell',r'}^*\,,
\qquad n\in\cN\,. \]
As $\canr,\lambda_\ell\in\DelNIo$ it follows
\[ a = \sum_{\ell=0}^n\sum_{r=1}^{m_\ell}
T_{\ell,r}aT_{\ell',r'}^* \,,\qquad a\in\CA(\Io') \,,\]
hence $T_{\ell,r}\in\CA(\Io')'=A(\Io)$. Thus we can define
$t_{\ell,r}=\pioi(T_{\ell,r})\in\NIo$, and we find
in particular
\[ \canr(n) = \sum_{\ell=0}^n\sum_{r=1}^{m_\ell}
t_{\ell,r} \, \lambda_\ell (n) \, t_{\ell',r'}^* \,,
\qquad n\in\NIo \,, \]
and this is in terms of sectors of $\NIo$
\be
[\canr] = \bigoplus_{\ell=0}^n \,\, m_\ell \, [\lambda_\ell] \,.
\labl{decom}

\section{$\alpha$-induction for nets of subfactors}
\lablsec{aind}

>From now on we assume that we have a given quantum field theoretical
net of subfactors $\cN\subset\cM$ over the index set $\Jz$, i.e.\
$N(I_1)\subset M(I_2)'$ if $I_1\cap I_2=\emptyset$. This implies
locality of the net $\cN$ but we even assume the net $\cM$ to be local,
and we also assume the net $\cA=\{ A(I)=\pio(N(I))\,,\,\,\,I\in\Jz \}$
to satisfy Haag duality. We also assume the net $\cN$ (or equivalently
the net $\cA$) to be strongly additive. 
Moreover, we require the net $\cN\subset\cM$
to be of finite index, $[M:N]<\infty$.  We fix an arbitrary interval
$\Io\in\Jz$ and take the corresponding endomorphism $\can$ of
Proposition \ref{LR1}.

\subsection{Definition of $\alpha$-induction}

In the following we set $\eps\lambda\mu=\epsp\lambda\mu$
for any $\lambda,\mu\in\DelNIo$. As usual, we denote by
$v\in\MIo$ and $w\in\NIo$ the isometries which intertwine
$\can\in\End(\cM)$ and its restriction $\canr\in\DelNIo$,
respectively, and satisfy $w^*v=[M:N]^{-1/2}\bfe=w^*\can(v)$.

\begin{lemma}
For $\lambda\in\DelNIo$ we have
\be
\Ad (\eps \lambda \canr) \circ \lambda \circ \can (v) =
\canr (\eps \lambda \canr ^*) \can (v) \,.
\ee
\lablth{AdlW}
\end{lemma}

\bproof
By the intertwining property of $v$ we find
$\can(v)^*\in\Hom_\NIo (\canr^2,\canr)$.
Hence we can apply the BFE, \erf {BFE3}, and obtain
\[ \eps \lambda \canr \cdot \lambda \circ \can (v)^*
= \can(v)^* \canr (\eps \lambda \canr)
\eps \lambda \canr \,, \]
hence
\[ \eps \lambda \canr \cdot \lambda \circ \can(v) \cdot
\eps \lambda \canr ^* = \canr (\eps \lambda \canr ^*)
\can(v) \,. \]
\eproof

If $I\in\Jz$ contains $\Io$ then
for $n\in\NI$ we have
$\Ad (\eps \lambda \canr ) \circ \lambda \circ \can (n) =
\Ad (\eps \lambda \canr ) \circ \lambda \circ \canr (n) =
\canr \circ \lambda (n)
\in \canr (\NI) \subset \can (\MI)$, and
note that then also
$\canr (\eps \lambda \canr ^*)\can(v) \in \can (\MI)$.
Since each $m\in\MI$ can be written as $m=nv$ for some $n\in\NI$
we find

\begin{corollary}
For any $I\in\Jz$ such that $\Io\subset I$ we have
\be
\Ad (\eps \lambda \canr) \circ \lambda \circ \can (\MI)
\subset \gamma (\MI) \,.
\ee
\lablth{well}
\end{corollary}
Now we are ready to define $\alpha$-induction --- just by
the formula (3.10) for the extended endomorphism in
Proposition 3.9 in \cite{lore}. However, we have shown that
this endomorphism leaves each algebra $\MI$ with $I\in\Jz$
such that $\Io\subset I$ invariant.

\begin{definition}
For $\lambda\in\DelNIo$ we define the $\alpha$-induced
endomorphism $\alpha_\lambda\in\End(\cM)$ by
\be
\alpha_\lambda = \cani \circ \Ad (\eps \lambda\canr)
\circ \lambda \circ \can \,.
\ee
\lablth{alpha}
\end{definition}
Thanks to Corollary \ref{well}, $\alpha_\lambda$ is well defined
and can also be considered as an element of $\End(\MI)$ as
long as $I\in\Jz$ contains $\Io$.
The definition of $\alpha$-induction is such that $\ala$ is an
extension of $\lambda$, i.e.\ we have
$\ala(n)=\lambda(n)$ obviously for $n\in\cN$.

\subsection{The main formula for $\alpha$-induction}

Choose $I_+\in\Jz$ such that $\Io<I_+$
and denote by $\can_+$ a (canonical) endomorphism associated to $I_+$
as in Proposition \ref{LR1}, and let $\canr_+$ its restriction
to $\cN$. Then the unitary $u=[M:N]\cdot E(v_+v^*)\in\cN$
intertwines $\can$ and $\can_+$ and relates isometries $v$ and
$v_+\in M(I_+)$ by $v_+=uv$ \cite{lore}. The proof of the following
lemma from \cite{lore} makes use of locality of the net $\cM$.

\begin{lemma}
We have
\be \eps \canr \canr v^2  = \eps \canr \canr ^*v^2= v^2\,, \qquad
\eps \canr \canr \can(v)  = \eps \canr \canr ^*\can(v)= \can(v) \,.
\ee
\lablth{eWW}
\end{lemma}

\bproof
By the intertwining property of $u$ we have in particular
$\canr_+=\Ad (u) \circ \canr$. Therefore $u=\ucp$ is a charge
transporter for $\canr$ and we can write
$\eps \canr\canr= u^* \canr(u)$. By locality of $\cM$ we find
$v_+v=vv_+$, i.e.\ $uvv=vuv=\canr(u)vv$, hence
$\eps \canr\canr v^2 \equiv u^*\canr (u) v^2 = v^2$. Since
$v^2=\can(v)v$ we obtain
$\eps \canr\canr \can(v)vv^* = \can(v)vv^*$ by right multiplication
with $v^*$. Application of the conditional expectation
yields $\eps \canr\canr \can(v) = \can(v)$ since
$E (vv^*) = w^*\can(vv^*)w=[M:N]^{-1}\,\bfe$.
Multiplying the obtained relations by $\eps \canr\canr ^*$ from
the left yields the full statement.
\eproof

Later we will use the following important

\begin{lemma}
Let $t\in\MIo$ such that
$t\lambda(n)=\mu(n)t$ for all $n\in\NIo$ and
some $\lambda,\mu\in\DelNIo$. Then
$t\in\Hom_\MIo(\ala,\amu)$.
\lablth{myst}
\end{lemma}

\bproof
As $\ala,\amu$ restrict, respectively, to $\lambda,\mu$
on $\NIo$ it suffices to show $t\ala(v)=\amu(v)t$.
Let $s=\can(t)$. Then clearly
$s\in\Hom_\NIo(\canr\circ\lambda,\canr\circ\mu)$.
By the BFE, \erf{BFE4}, we obtain
\[ \eps {\canr\circ\mu} \canr \,s = \canr (s) \eps \canr \canr
\canr (\eps \lambda \canr) \,.\]
Since
$\eps {\canr\circ\mu} \canr = \eps \canr\canr \canr(\eps \mu\canr)$
we find
\[ s \, \canr (\eps \lambda\canr ^*) = \canr (\eps \mu\canr ^*)
\eps \canr\canr ^* \canr (s) \eps \canr\canr \,.\]
So let us compute
\[ \bearll
s \cdot \Ad (\eps \lambda \canr ) \circ \lambda \circ \can (v)
&= s \, \canr (\eps \lambda\canr ^*) \can(v) \\[.4em]
&= \canr (\eps \mu\canr ^*)
\eps \canr\canr ^* \canr (s) \eps \canr\canr \can(v) \\[.4em]
&= \canr (\eps \mu\canr ^*)
\eps \canr\canr ^* \canr (s)  \can(v) \\[.4em]
&= \canr (\eps \mu\canr ^*) \eps \canr\canr ^*  \can(v)s \\[.4em]
&= \canr (\eps \mu\canr ^*)  \can(v)s \\[.4em]
&= \Ad (\eps \mu \canr ) \circ \mu \circ \can (v) \cdot s \,,
\eear\]
where we repeatedly used Lemmata \ref{AdlW}, \ref{eWW}, and also that
$\canr(s)\can(v) = \can^2(t)\can(v) = \can(v)\can(t) = \can(v)s$.
Thanks to Corollary \ref{well} we can now apply $\cani$ and
obtain $t\ala(v)=\amu(v)t$.
\eproof

Note that we obtained Lemma \ref{myst} just by the following
ingredients: Haag duality and strong additivity of the net $\cA$,
implying existence of statistics operators and the BFEs for local
intertwiners of endomorphisms in $\DelNIo$,
and locality of the net $\cM$, implying
Lemma \ref{eWW}, and of course, finiteness of the index guaranteeing
the existence of the isometry $v$. Now consider the following
special situation $\lambda=\mu=\id$ in Lemma \ref{myst}. First note
that $\alpha_\id=\id$ by the definition of $\alpha$-induction.
Then for each $t\in\NIo'\cap\MIo$ we find
$t\in\Hom_\MIo(\id,\id)=\MIo'\cap\MIo$, i.e.\
\[ \NIo'\cap\MIo \subset \MIo'\cap\MIo = \bbC \,\bfe \,, \]
and $\Io\in\Jz$ was arbitrary. Somewhat surprisingly, we gained

\begin{corollary}
Let $\cN\subset\cM$ be a directed quantum field theoretical net
of subfactors over $\Jz$ with finite index. If $\cN$ is strongly
additive and has a Haag dual
subrepresentation and $\cM$ satisfies locality,
then $\cN\subset\cM$ is a net of irreducible subfactors.
\lablth{surp}
\end{corollary}

Another immediate consequence of Lemma \ref{myst} is the following

\begin{corollary}
If $[\lambda]=[\mu]$ for some $\lambda,\mu\in\DelNIo$,
then $[\ala]=[\amu]$.
\lablth{secpres}
\end{corollary}
(Here and in the following we use the sector brackets for sectors
of either $\NIo$ or $\MIo$.)

\begin{lemma}
If $n\in\cN$ then $nv=0$ implies $n=0$. Similarly, for $m\in\cM$,
$w^*\can(m)=0$ implies $m=0$.
\lablth{inj}
\end{lemma}

\bproof
This follows from the identities
$n=[M:N]^\h nw^*\can(v)=[M:N]^\h w^*\can(nv)$, $n\in\cN$,
and $m=[M:N]^\h w^*vm=[M:N]^\h w^*\can(m)v$, $m\in\cM$.
\eproof

We are now ready to prove the main formula for $\alpha$-induction
given in the following

\begin{theorem}
For $\lambda,\mu\in\DelNIo$ we have
\be
\la \alpha_\lambda , \amu \ra_\MIo =
\la \canr \circ \lambda , \mu \ra_\NIo \,.
\ee
\lablth{main}
\end{theorem}

\bproof
We first show ``$\le$". Let $t\in\Hom_\MIo(\alpha_\lambda,\amu)$.
We show that $r=w^*\can(t)\in\Hom_\NIo(\canr\circ\lambda,\mu)$.
Clearly, $r\in\NIo$. By assumption, we have
$t\alpha_\lambda(m)=\amu(m)t$ for all $m\in\MIo$.
Restriction to $\NIo$ and application of $\can$ yields
$\can(t)\in\Hom_\NIo(\canr\circ\lambda,\canr\circ\mu)$.
It follows for all $n\in\NIo$
\[ 
r \cdot \canr \circ \lambda (n) = w^* \cdot\can(t) \cdot
\canr \circ \lambda (n) 
= w^* \cdot \canr\circ\mu(n)\cdot\can(t) = \mu(n) \, r \]
since $w^*\canr(n) = nw^*$. By Lemma \ref{inj}
the map $t\mapsto r=w^*\can(t)$ is injective, thus
``$\le$" is proven.

We now turn to ``$\ge$". Suppose
$r\in\Hom_\NIo(\canr\circ\lambda,\mu)$
is given. We show that $t=rv\in\Hom_\MIo(\ala,\amu)$.
Clearly, $t=rv\in\MIo$, and we have for all $n\in\NIo$
\[ t\,\lambda (n) = rv\, \lambda(n) =
r \cdot \canr\circ\lambda(n)\cdot v
=\mu (n)\, rv = \mu(n)\, t \,. \]
Hence, by Lemma \ref{myst}, we have
$t\in\Hom_\MIo(\ala,\amu)$.
By Lemma \ref{inj}, the map $r\mapsto t=rv$ is injective;
the proof is complete.
\eproof

\subsection{Homomorphism property of $\alpha$-induction}
\lablsec{ahom}

As $\ala$ restricts to $\lambda$ on $\NIo$ which
is of finite index in $\MIo$, we find $d_{\ala}=d_\lambda$.
This is an immediate consequence of
the multiplicativity of the minimal index \cite{lon3}: Consider
the chain of inclusions $\ala(\NIo)\subset\NIo\subset\MIo$. Choose
$\eta\in\End(\MIo)$ such that $\eta(\MIo)=\NIo$. Then
$[\MIo:\NIo]=d_\eta^2$ and
$[\MIo:\ala(\NIo)]=d_{\ala}^2 d_\eta^2$, hence
$[\NIo:\ala(\NIo)]=d_{\ala}^2$ but
$[\NIo:\ala(\NIo)]\equiv[\NIo:\lambda(\NIo)]
=d_\lambda^2$, thus indeed $d_{\ala}=d_\lambda$.
However, there are more properties.

\begin{lemma}
For any $\lambda,\mu\in\DelNIo$ we have
$\alpha_{\lambda\circ\mu}=\alpha_\lambda\circ\amu$.
\lablth{mult}
\end{lemma}

\bproof
We compute
\[ \bearll
\alpha_{\lambda\circ\mu} &= \cani\circ
\Ad(\eps{\lambda\circ\mu}\canr)\circ\lambda\circ
\mu\circ\can\\[.4em]
&= \cani\circ\Ad(\eps\lambda\canr
\lambda(\eps\mu\canr))\circ\lambda\circ
\mu\circ\can\\[.4em]
&= \cani\circ\Ad(\eps\lambda\canr)\circ\lambda\circ
\Ad(\eps\mu\canr)\circ\mu\circ\can\\[.4em]
&= \alpha_\lambda\circ\amu \,, \eear \]
where we used \erf{fuenf}.
\eproof

As $\eps\lambda\mu\in\Hom_\NIo(\lambda\circ\mu,\mu\circ\lambda)$
we obtain from Lemma \ref{myst} that
$\eps\lambda\mu\alpha_{\lambda\circ\mu}(m)=
\alpha_{\mu\circ\lambda}(m)\eps\lambda\mu$ for all $m\in\MIo$,
in particular for $m=v$. Since $\cM=\cN v$ we obtain from
Lemma \ref{mult} the following

\begin{corollary}
For $\lambda,\mu\in\DelNIo$ we have
\be
\amu\circ\ala = \Ad (\eps\lambda\mu)\circ\ala\circ\amu \,.
\ee
\lablth{abraid}
\end{corollary}
As $\ala$ restricts to $\lambda$ on $\cN$ we clearly have
$\ala(\eps\mu\nu)=\lambda(\eps\mu\nu)$ for
$\lambda,\mu,\nu\in\DelNIo$. Therefore, by rewriting the YBE,
\erf{YBEe}, and recalling that
$\eps\lambda\lambda\in\ala^2(\MIo)'\cap \MIo$
by Corollary \ref{abraid}, we arrive at

\begin{corollary}
For $\lambda,\mu,\nu\in\DelNIo$ we have the YBE
\be
\alpha_\nu(\eps\lambda\mu) \, \eps\lambda\nu \, \ala(\eps\mu\nu) =
\eps\mu\nu \, \amu(\eps\lambda\nu) \, \eps\lambda\mu \,,
\labl{YBEea}
in particular, the endomorphisms $\ala$ are braided endomorphisms,
i.e.\ setting $\sigma_i=\ala^{i-1}(\eps\lambda\lambda)$,
$i=1,2,3,\ldots$, yields a representation of the braid group $B_\infty$.
\end{corollary}

Next we show that $\alpha$-induction preserves also
sums of sectors.

\begin{lemma}
Let $\lambda,\lambda_1,\lambda_2\in\DelNIo$ such that
$[\lambda]=[\lambda_1]\oplus[\lambda_2]$. Then
$[\alpha_\lambda]=[\alpha_{\lambda_1}]\oplus[\alpha_{\lambda_2}]$.
\lablth{sums}
\end{lemma}

\bproof
As $[\lambda]=[\lambda_1]\oplus[\lambda_2]$
we have isometries $y_1,y_2\in\NIo$ fulfilling the
relations of $\cO_2$, $y_i^*y_j=\del ij \, \bfe$,
$\sum_{i=1}^2 y_iy_i^* = \bfe$,
and
\[ \lambda (n) = \sum_{i=1}^2 y_i \, \lambda_i (n) \, y_i^*
\,,\qquad n\in\NIo \,.\]
We now choose an interval $I_+\in\Jz$ such that
$\Io<I_+$. Note that
$y_i\in\Hom_\NIo(\lambda_i,\lambda)=\Hom_\cN(\lambda_i,\lambda)$,
$i=1,2$. Choose a charge transporter $\ucp\in\cN$ such that
$\canr_+=\Ad(\ucp)\circ\canr\in\Delta_\cN(I_+)$.
Then we have
\[ \eps \lambda\canr = \ucp^* \lambda (\ucp) =
\sum_{i=1}^2 \ucp^* \lambda(\ucp) \, y_i y_i^*
= \sum_{i=1}^2 \ucp^* \, y_i \,
\lambda_i (\ucp) \, y_i^* \,.\]
Since $y_i\in\NIo$ we also find $\canr_+(y_i)=y_i$, $i=1,2$,
and thus we compute for $n\in\NIo$
\[ \bearll
\Ad (\eps \lambda\canr)\circ  \lambda (n)
&=  \sum_{i=1}^2 \ucp^* \, y_i \, \lambda_i (\ucp \, n \, \ucp^*)
\, y_i^* \ucp \\[.4em]
&= \sum_{i=1}^2 \ucp^* \, \canr_+ (y_i) \, \lambda_i (\ucp \,n\,\ucp^*)
\, \canr_+ (y_i^*) \, \ucp \\[.4em]
&= \sum_{i=1}^2 \canr (y_i) \, \ucp^* \ \lambda_i (\ucp \,n\,\ucp^*)
\, \ucp \,\canr (y_i^*) \, \\[.4em]
&= \sum_{i=1}^2 \canr (y_i) \cdot \Ad (\eps{\lambda_i}\canr)
\circ \lambda_i (n) \cdot \canr (y_i^*) \,. \eear \]
Specializing to $n=\can(m)$, $m\in\MIo$, and applying $\cani$ yields
\[ \alpha_\lambda(m) = \sum_{i=1}^2 y_i \, \alpha_{\lambda_i}(m) \, y_i^*
\,, \qquad m\in\MIo \,,\]
the lemma is proven.
\eproof

For sectors with finite statistical dimension we can show that
$\alpha$-induction preserves also sector conjugation.

\begin{lemma}
If $\lambdab\in\DelNIo$ is a conjugate to $\lambda\in\DelNIo$,
$d_\lambda<\infty$, then $\alab$ is a conjugate to $\ala$, i.e.\
$[\alab]=[\bala]$.
\lablth{conj}
\end{lemma}

\bproof
Using Lemma \ref{mult}, Theorem \ref{main} and \erf{frod}
we get
\[ \bearll
\la \ala,\ala \ra_\MIo &= \la \canr\circ\lambda,\lambda \ra_\NIo \\[.4em]
&= \la \canr\circ\lambda\circ\lambdab,\id_\NIo \ra_\NIo \\[.4em]
&= \la \alpha_{\lambda\circ\lambdab},\id_\MIo \ra_\MIo \\[.4em]
&= \la \ala\circ\alab,\id_\MIo \ra_\MIo \\[.4em]
&= \la \ala, \balab \ra_\MIo \,. \eear \]
Replacing $\lambda$ by $\lambdab$ yields
$\la \alab,\alab \ra_\MIo = \la \alab,\bala \ra_\MIo$ whereas
conjugation yields $\la \bala,\bala \ra_\MIo = \la \bala,\alab \ra_\MIo$.
Thus we found
\[ \la \alab,\alab \ra_\MIo = \la \alab,\bala \ra_\MIo
= \la \bala,\bala \ra_\MIo \,, \]
and because we assumed finite statistical dimensions,
these expressions are finite. Then this implies the statement.
\eproof

Next we want to discuss certain commutativity rules between
sectors arising from $\alpha$-induction.

\begin{lemma}
Let $\lambda,\mu,\rho\in\DelNIo$ and $r\in\MIo$ such that
$r\lambda(n)=\mu(n)r$ for all $n\in\NIo$. Then we have
$r \eps \rho\lambda = \eps \rho\mu \alpha_\rho(r)$.
\lablth{voranat}
\end{lemma}

\bproof
Note that $s=\can(r)\in\Hom_\NIo(\canr\circ\lambda,\canr\circ\mu)$.
Thus the BFE, \erf{BFE3}, yields
\[ s\, \canr(\eps\rho\lambda) \eps \rho\canr =
\eps \rho {\canr\circ\mu} \, \rho(s) \,, \]
hence we obtain by using \erf{sechsn}
\[ s\, \canr(\eps\rho\lambda) = \canr(\eps\rho\mu) \, \eps\rho\canr
\rho(s) \eps\rho\canr ^* \,, \]
and applying $\cani$ yields the statement.
\eproof

\begin{proposition}
Let $\lambda,\mu\in\DelNIo$ and $\beta\in\End(\MIo)$ such that
$[\beta]$ is a subsector of $[\amu]$. Then
$[\ala\circ\beta]=[\beta\circ\ala]$.
\lablth{abel}
\end{proposition}

\bproof
By assumption, there is an isometry $t\in\MIo$, $t^*t=\bfe$,
such that
\[ t\,\beta(m) = \amu (m)\, t \,, \qquad m\in\MIo \,.\]
Then $u=t^* \eps\lambda\mu \ala(t)\in
\Hom_{\MIo}(\ala\circ\beta,\beta\circ\ala)$ as we have for
all $m\in\MIo$
\[ \bearll
t^* \eps\lambda\mu \ala (t) \cdot \ala\circ\beta (m)
&= t^* \eps\lambda\mu \cdot \ala \circ \amu (m)
\cdot \ala(t) \\[.4em]
&= t^* \cdot \amu\circ\ala (m) \cdot \eps\lambda\mu
\ala (t) \\[.4em]
&= \beta\circ\ala(m) \cdot t^* \eps\lambda\mu \ala (t) \,,
\eear \]
where we used Corollary \ref{abraid}. All we have to show is
that $u$ is unitary. Note that $tt^*\in\Hom_\MIo(\amu,\amu)$
and hence in particular $tt^*\in\mu(\NIo)'\cap\MIo$ as $\amu$
restricts to $\mu$ on $\NIo$. Then Lemma \ref{voranat}
yields $tt^*\eps\lambda\mu=\eps\lambda\mu\ala(tt^*)$.
Therefore
\[ u^*u= \ala(t^*) \eps\lambda\mu ^* tt^* \eps\lambda\mu \ala(t)
= \ala(t^*tt^*t)=\bfe \,, \]
and
\[ uu^* = t^* \eps\lambda\mu \ala(tt^*) \eps\lambda\mu ^*t
=t^*tt^*t = \bfe \,, \]
the proof is complete.
\eproof

\subsection{$\sigma$-restriction and $\alpha\sigma$-reciprocity}
\lablsec{sres}

In \cite{lore} there is also defined a restriction for endomorphisms.
In our context, we will call that $\sigma$-restriction.

\begin{definition}
For $\beta\in\End(\cM)$ the $\sigma$-restricted endomorphism
$\sib\in\End(\cN)$ is defined by
\be
\sib = \can \circ \beta |_\cN \,.
\ee
\end{definition}

If $\beta\in\End(\cM)$ leaves $\MI$ invariant for $I\in\Jz$,
$\Io\subset I$, then clearly $\sib$ leaves $\NI$ invariant.
Moreover, the formula $\sib(n)=\can\circ\beta(n)$, $n\in\NI$,
defines also a map from $\End(\MI)$ to $\End(\NI)$.
For $\lambda\in\DelNIo$ we obviously have
$\sigma_{\ala}=\canr\circ\lambda$ so that in particular
$[\lambda]$ is a subsector of $[\sigma_{\ala}]$. It is
natural to ask whether $[\beta]$ is a subsector of
$[\alpha_\sib]$. For localized, transportable $\beta$
we are going to prove an even stronger result which is a sort of
Frobenius reciprocity for $\alpha$-induction and $\sigma$-restriction.
For this we need some more preparation.

Clearly, if $\beta$ is localized in $\Io$ then so is $\sib$
as for $n\in\cC_\cN(\Io')$ we find
$\sib(n)=\can\circ\beta(n)=\can(n)=\canr(n)=n$ since
$\canr$ is localized in $\Io$. Now suppose that $\beta$ is
also transportable: For each $I_1\in\Jz$
we have unitary charge transporters $\QbI\in\cM$ such that
$\beta_{I_1}=\Ad(\QbI)\circ\beta$ is localized in $I_1$.

\begin{lemma}
If $\beta\in\DelMIo$ then $\sib\in\DelNIo$.
Namely, for any $I_1\in\Jz$ we have
$\sigma_{\beta,I_1}=\Ad(u_{\sib;\Io,I_1})\circ\sib\in\Delta_\cN(I_1)$
with
\be
u_{\sib;\Io,I_1} = \ucI \, \can (\QbI) \,.
\ee
\lablth{sibdel}
\end{lemma}

\bproof
We have to show that
$\sigma_{\beta,I_1}=\Ad(u_{\sib;\Io,I_1})\circ\sib$ is localized
in $I_1$. Now for $n\in\CN(I_1')$ we have
\[ \bearll
\sigma_{\beta,I_1}(n) &= \ucI \, \can (\QbI) \cdot
\can \circ \beta(n) \cdot \can (\QbI)^* \, \ucI^* \\[.4em]
&= \ucI \cdot \can \circ \beta_{I_1} (n) \cdot \ucI^* \\[.4em]
&= \ucI \, \can (n) \, \ucI^* \\[.4em]
&= \ucI \, \canr (n) \, \ucI^* \\[.4em]
&= \canr_{I_1} (n) =n \,,
\eear \]
since $\canr_{I_1}=\Ad(\ucI)\circ\canr$ is localized in $I_1$.
\eproof

For some interval $I_-\in\Jz$ such that $I_-<\Io$
we set $\Qbm = Q_{\beta;\Io,I_-}$.

\begin{lemma}
For $\beta\in\DelMIo$ we have
\be
\eps \sib\canr = \can^2(\Qbm)^* \, \eps \canr\canr \,
\can(\Qbm) \,.
\ee
\lablth{epssib}
\end{lemma}

\bproof
We compute
\[ \bearll
\eps \sib\canr &= \epsm \canr\sib ^*
= \canr (u_{\sib;\Io,I_-})^*u_{\sib;\Io,I_-} \\[.4em]
&= \canr (\can (\Qbm)^* \ucm^*) \, \ucm \can (\Qbm) \\[.4em]
&= \can^2 (\Qbm)^* \canr(\ucm)^* \, \ucm \can (\Qbm) \\[.4em]
&= \can^2 (\Qbm)^* \epsm \canr\canr ^* \, \can (\Qbm) \\[.4em]
&= \can^2 (\Qbm)^* \eps \canr\canr \, \can (\Qbm) \,,
\eear \]
where we used \erf{viern}.
\eproof

For $I\in\Jz$ let $\Delta_\cM^{(0)}(I)$ denote the set of transportable
endomorphisms localized in $I$ which leave $M(K)$ invariant for any
$K\in\Jz$ with $I\subset K$. Note that $\lambda(M(K))\subset M(K)$
for $\lambda\in\Delta_\cM(I)$ is automatically satisfied if $\cM$
is Haag dual, i.e.\ $\Delta_\cM^{(0)}(I)=\Delta_\cM(I)$ in this
case. However, in order to be as general as possible we do
not assume Haag duality of $\cM$ (although it is satisfied in the
applications we have in mind) but we do need invariance of local
algebras as we often consider elements of $\DelMIO$ as elements of
$\End(M(K))$ for $I\subset K$.

\begin{lemma}
Let $t\in\MIo$ such that
$t\lambda(n)=\beta(n)t$ for all $n\in\NIo$ and
some $\lambda\in\DelNIo$ and $\beta\in\DelMIoO$.
Then $t\in\Hom_\MIo(\ala,\beta)$.
\lablth{myst'}
\end{lemma}

\bproof
As $\ala(n)=\lambda(n)$ for all $n\in\NIo$ it suffices to show
$t\ala(v)=\beta(v)t$. Let $s=\can(t)$. Then clearly
$s\in\Hom_\NIo(\canr\circ\lambda,\sib)$.
By the BFE, \erf{BFE4}, we obtain
\[ s \, \canr (\eps \lambda\canr ^*) = \eps \sib\canr ^* \canr (s)
\eps \canr \canr \,.\]
So let us compute
\[ \bearll
s \cdot \Ad (\eps \lambda \canr ) \circ \lambda \circ \can (v)
&= s \, \canr (\eps \lambda\canr ^*) \can(v) \\[.4em]
&= \eps \sib\canr ^* \canr (s) \eps \canr\canr \can(v) \\[.4em]
&= \eps \sib\canr ^* \canr (s) \can(v) \\[.4em]
&= \eps \sib\canr ^* \can(v)s \\[.4em]
&= \can (\Qbm)^* \eps \canr\canr ^* \can^2 (\Qbm) \can(v)s \\[.4em]
&= \can (\Qbm)^* \eps \canr\canr ^* \can(v) \can (\Qbm) s \\[.4em]
&= \can (\Qbm)^* \can(v) \can (\Qbm) s \\[.4em]
&= \can (\Qbm^* v \Qbm) s \\[.4em]
&= \can (\Qbm^* \beta_{I_-} (v) \Qbm) s \\[.4em]
&= \can \circ \beta (v) \cdot s \,,
\eear\]
where we repeatedly used Lemmata \ref{AdlW}, \ref{eWW}
and \ref{epssib}. Applying $\cani$ yields
$t\ala(v)=\beta(v)t$.
\eproof

Now we are ready to prove the reciprocity theorem.

\begin{theorem}
For $\lambda\in\DelNIo$ and $\beta\in\DelMIoO$ we have
$\alpha\sigma$-reciprocity,
\be
\la \ala, \beta \ra_\MIo = \la \lambda, \sib \ra_\NIo \,.
\ee
\lablth{asrep}
\end{theorem}

\bproof
We first show ``$\le$". Let $t\in\Hom_\MIo(\ala,\beta)$.
We show that $r=\can(t)w\in\Hom_\NIo(\lambda,\sib)$.
Clearly, $r\in\NIo$. By assumption, we have
$t\alpha_\lambda(m)=\beta(m)t$ for all $m\in\MIo$.
Restriction to $\NIo$ and application of $\can$ yields
$\can(t)\in\Hom_\NIo(\canr\circ\lambda,\sib)$.
It follows for all $n\in\NIo$
\[ r \, \lambda (n) = \can(t)w \, \lambda (n) 
= \can(t) \cdot \canr \circ \lambda (n) \cdot w
= \sib (n) \, \can(t)w = \sib (n) \, r \,. \]
By Lemma \ref{inj} the map $t\mapsto r=\can(t)w$
is injective, thus ``$\le$" is proven.

We now turn to ``$\ge$". Suppose $r\in\Hom_\NIo(\lambda,\sib)$
is given. We show that $t=v^*r\in\Hom_\MIo(\ala,\beta)$.
Clearly, $t=v^*r\in\MIo$, and we have for all $n\in\NIo$
\[ t\,\lambda (n) = v^*r\, \lambda(n) =
v^*\,\sib(n)\,r=v^* \cdot \can\circ\beta(n)\cdot r
=\beta (n)\, v^*r = \beta(n)\, t \,. \]
Hence, by Lemma \ref{myst'}, we have
$t\in\Hom_\MIo(\ala,\beta)$.
It follows again from Lemma \ref{inj} that
the map $r\mapsto t=v^*r$ is injective; the proof is complete.
\eproof

It follows from the proof that we have
$v^*\in\Hom_\MIo(\alpha_{\sib},\beta)$ since
$\bfe\in\Hom_\NIo(\sib,\sib)$. (Recall
$\sib\in\DelNIo$ by Lemma \ref{sibdel}.) We conclude that
$[\beta]$ is a subsector of $[\alpha_{\sib}]$.

{\it Remark.} Note that Theorem \ref{asrep} is {\em not} a
generalization of Theorem \ref{main} since we assumed in
particular that $\beta$ is localized. However, $\amu$
is in general not localized; it is localized if and only
if the monodromy $\eps \mu\canr \eps \canr\mu$ is trivial
(Prop.\ 3.9 in \cite{lore}).

Note that $\sigma$-restriction does not preserve sector
products, i.e.\ $[\sigma_{\beta_1}\circ\sigma_{\beta_2}]$
is in general different from $[\sigma_{\beta_1\circ\beta_2}]$,
e.g.\ for $\beta_1=\beta_2=\id$. However, we add the following

\begin{lemma}
Let $\beta,\beta_1,\beta_2\in\End(\MIo)$.
If $[\beta]=[\beta_1]\oplus[\beta_2]$ then
$[\sib]=[\sigma_{\beta_1}]\oplus[\sigma_{\beta_2}]$.
If $[\beta_1]=[\beta_2]$ then
$[\sigma_{\beta_1}]=[\sigma_{\beta_2}]$.
\lablth{sihom}
\end{lemma}

\bproof
If $[\beta]=[\beta_1]\oplus[\beta_2]$ then there are
isometries $t_1,t_2\in\MIo$ satisfying the relations of $\cO_2$ and
$\beta(m)=\sum_{i=1}^2 t_i\beta_i(m) t_i^*$ for $m\in\MIo$.
Then $s_i=\can(t_i)$ satisfy the relations of $\cO_2$ as well and
\[ \sib(n) = \can\circ\beta(n) = \sum_{i=1}^2
s_i\cdot\can\circ\beta_i(n)\cdot s_i^* = \sum_{i=1}^2
s_i \sigma_{\beta_i}(n) s_i^*\,, \qquad n\in\NIo\,. \]
If $[\beta_1]=[\beta_2]$ then $\beta_2=\Ad(u)\circ\beta_1$
with some unitary $u\in\MIo$. Then clearly
$\sigma_{\beta_2}=\Ad(\can(u)) \circ \sigma_{\beta_1}$,
and $\can(u)\in\NIo$ is unitary.
\eproof

\subsection{The inverse braiding}
\lablsec{obra}

We have used the statistics operators
$\eps \lambda\canr \equiv \epsp \lambda\canr$ for the definition
of the $\alpha$-induced endomorphism $\ala\equiv\alap$.
Of course, all our results we derived hold similarly for the
endomorphims $\alam$, analogously defined by use of
$\epsm \lambda\canr$. However, $\ala$ and $\alam$
are in general not the same. In this subsection we investigate
several relations between $\ala$ and $\alam$. The following
Proposition is instructive.

\begin{proposition}
For $\lambda\in\DelNIo$ the following are equivalent:
\begin{enumerate}
\item $[\ala]=[\alam]$,
\item $\ala=\alam$,
\item The monodromy is trivial:
$\eps\lambda\canr\eps\canr\lambda=\bfe$.
\end{enumerate}
\end{proposition}

\bproof
If $[\ala]=[\alam]$ then there is a unitary
$u\in\Hom_\MIo(\ala,\alam)$, i.e.
$u\ala(m)=\alam(m)u$ for all $m\in\MIo$. Restriction yields
$u\lambda(n)=\lambda(n)u$ for all $n\in\NIo$. By Lemma \ref{myst}
we find $u\in\Hom_\MIo(\ala,\ala)$, in particular
$u\ala(v)=\ala(v)u$, hence
$\ala(v)u=\alam(v)u$, thus $\ala(v)=\alam(v)$. But
$\ala(n)=\lambda(n)=\alam(n)$ for all $n\in\cN$, therefore
$\ala(m)=\alam(m)$ for any $m\in\cM$,
proving $\ala=\alam$. Now by Lemma \ref{AdlW} we have
$\ala(v)=\eps \lambda\canr ^* v$, and similarly
$\alam(v)=\epsm \lambda\canr ^* v = \eps \canr\lambda v$. Therefore
$\ala(v)=\alam(v)$ implies $\eps\lambda\canr\eps\canr\lambda v=v$
and hence $\eps\lambda\canr\eps\canr\lambda=\bfe$ by Lemma \ref{inj}.
Now if the monodromy is trivial then
$\eps \lambda\canr=\epsm \lambda\canr$, and this trivially leads
to $[\ala]=[\alam]$.
\eproof

Nevertheless we have the following

\begin{lemma}
For $\lambda,\mu\in\DelNIo$ we have
\be
\Ad (\eps \lambda\mu) \circ \ala \circ \amum = \amum \circ \ala \,.
\ee
\lablth{oblm}
\end{lemma}

\bproof
As $\ala$ and $\amum$ restrict to $\lambda$ and $\mu$, respectively,
on $\cN$ it suffices to show
\[ \eps \lambda\mu \cdot \ala\circ\amum(v) =
\amum \circ \ala (v) \cdot \eps \lambda\mu \,. \]
Recall $\ala(v)=\eps \lambda\canr ^* v$ by Lemma \ref{AdlW}, and
similarly $\amum(v)=\epsm \mu\canr ^* v = \eps \canr\mu v$.
The YBE, \erf{YBEe}, can be written as
\[ \eps \lambda\mu \lambda( \eps\canr\mu) \eps \lambda\canr ^* =
\mu (\eps \lambda\canr ^*) \eps \canr\mu \canr(\eps \lambda\mu) \,.\]
Now we compute
\[ \bearll
\eps \lambda\mu \cdot \ala\circ\amum (v)
&= \eps \lambda\mu \ala(\eps \canr\mu v) \\[.4em]
&= \eps \lambda\mu \lambda (\eps \canr\mu)
   \eps \lambda\canr ^* v \\[.4em]
&= \mu (\eps \lambda\canr ^*) \eps \canr\mu
   \canr(\eps \lambda\mu) v \\[.4em]
&= \mu (\eps \lambda\canr ^*) \eps \canr\mu \,
   v \, \eps \lambda\mu  \\[.4em]
&= \amum (\eps \lambda\canr ^* v) \eps \lambda\mu \\[.4em]
&= \amum \circ \ala (v) \cdot \eps \lambda\mu \,,
\eear \]
proving the lemma.
\eproof

The following lemma establishes a sort of naturality equations
for the $\alpha$-induced endomorphisms.

\begin{lemma}
Let $\lambda,\mu,\rho\in\DelNIo$. For an $r\in\MIo$ such that
$r\lambda(n)=\mu(n)r$ for all $n\in\NIo$ we have
\bea
\alpha_\rho^\pm (r) \, \epsmp \lambda\rho
&=&\, \epsmp \mu\rho \,  r \,, \label{anat2}\\
r \, \epspm \rho\lambda
&=& \epspm \rho\mu \, \alpha_\rho^\pm (r) \,. \label{anat1}
\eea
\lablth{anat}
\end{lemma}

\bproof
Completely analogous to Lemma \ref{voranat} we also obtain
$r\epsm\rho\lambda=\epsm\rho\mu \alpha_\rho^-(r)$, establishing
\erf{anat1}. Now note that $r^*\mu(n)=\lambda(n)r^*$ for all
$n\in\NIo$, therefore we can apply \erf{anat1} yielding
\erf{anat2} by use of \erf{viern}.
\eproof

We are now ready to prove the following

\begin{proposition}
Let $\lambda,\mu\in\DelNIo$ and $\beta,\delta\in\End(\MIo)$
such that $[\beta]$ and $[\delta]$ are subsectors of
$[\ala]$ and $[\amum]$, respectively. Then
$[\beta\circ\delta]=[\delta\circ\beta]$.
\lablth{obabel}
\end{proposition}

\bproof
By assumption, there are isometries $t,s\in\MIo$, $t^*t=s^*s=\bfe$,
such that
\[ t\,\beta(m) = \ala (m)\, t \,, \qquad
s\,\delta(m)=\amum(m)\,s \,, \qquad m\in\MIo \,.\]
Then $u=s^*\amum(t^*) \eps\lambda\mu \ala(s)t \in
\Hom_{\MIo}(\beta\circ\delta,\delta\circ\beta)$ as we have for
all $m\in\MIo$
\[ \bearll
s^*\amum(t^*) \eps\lambda\mu \ala(s)t \cdot \beta\circ\delta (m)
&=s^*\amum(t^*) \eps\lambda\mu \ala(s) \cdot \ala \circ
  \delta (m) \cdot t \\[.4em]
&=s^*\amum(t^*) \eps\lambda\mu \cdot \ala\circ\amum (m)
  \cdot \ala (s)t \\[.4em]
&= s^*\amum(t^*) \cdot \amum\circ\ala(m) \cdot \eps\lambda\mu
  \ala(s)t \\[.4em]
&= s^* \cdot \amum\circ\beta(m) \cdot \amum(t^*)\eps\lambda\mu
  \ala(s)t \\[.4em]
&= \delta\circ\beta \cdot s^*\amum(t^*) \eps\lambda\mu \ala(s)t \,,
\eear \]
where we used Lemma \ref{oblm}. All we have to show is
that $u$ is unitary. Note that $tt^*\in\Hom_\MIo(\ala,\ala)$
and $ss^*\in\Hom_\MIo(\amum,\amum)$ and hence in particular
$tt^*\in\lambda(\NIo)'\cap\MIo$ and
$ss^*\in\mu(\NIo)'\cap\MIo$ as $\ala$ and $\amum$
restrict to $\lambda$ and $\mu$, respectively, on $\NIo$.
Then Lemma \ref{anat} yields
$\amum(tt^*)\eps\lambda\mu=\eps\lambda\mu tt^*$ by \erf{anat2}
and $ss^*\eps\lambda\mu=\eps\lambda\mu\ala(ss^*)$ by \erf{anat1}.
Therefore
\[ \bearll
u^*u &= t^* \ala(s^*) \eps\lambda\mu ^* \amum(t) ss^* \amum(t^*)
  \eps\lambda\mu \ala(s)t \\[.4em]
&= t^* \ala(s^*) \eps\lambda\mu ^* ss^* \amum(tt^*)
  \eps\lambda\mu \ala(s)t \\[.4em]
&= t^* \ala(s^*) \eps\lambda\mu ^* ss^* \eps\lambda\mu 
  tt^* \ala(s)t \\[.4em]
&= t^* \ala(s^*ss^*s) tt^*t = \bfe \,,
\eear \]
and
\[ \bearll
uu^* &= s^* \amum(t^*) \eps\lambda\mu \ala(s) tt^* \ala(s^*)
  \eps\lambda\mu ^* \amum(t) s \\[.4em]
&=  s^* \amum(t^*) \eps\lambda\mu tt^* \ala(ss^*)
  \eps\lambda\mu ^* \amum(t) s \\[.4em]
&= s^* \amum(t^*tt^*) ss^* \amum(t)s = \bfe \,,
\eear \]
the proof is complete.
\eproof

Recall that $[\beta]$ is a subsector of $[\alpha_{\sib}]$ for any
$\beta\in\DelMIoO$, and in the same way it is a subsector of
$[\alpha^-_{\sib}]$. From Proposition \ref{obabel} we obtain
immediately

\begin{corollary}
For any $\beta\in\DelMIoO$ and any $\delta\in\End(\MIo)$
such that $[\delta]$ is a subsector of some $[\amu]$,
$\mu\in\DelNIo$, we have $[\beta\circ\delta]=[\delta\circ\beta]$.
\lablth{markabel}
\end{corollary}

\section{Miscellanea}

\subsection{The results in terms of sector algebras}

We now want to present our results in the language of
sector algebras. We need some preparation.

\begin{definition}
Let $V$ be a (real or complex), finite dimensional,
unital, associative algebra
(with addition $\oplus$ and multiplication $\times$)
together with a basis $\cV=\{v_0,v_1,v_2,\ldots,v_{d-1}\}$
(in the linear space sense) such that $\bfe\in\cV$, say
$v_0=\bfe$. Let $\N ijk$ be the structure constants,
defined by
\be
v_i\times v_j=\bigoplus_{k=0}^{d-1} \N ijk\, v_k \,.
\ee
If
\begin{enumerate}
\item (Conjugation) there is an involutive permutation
$i \mapsto \co i$, $\co {\co i}=i$, satisfying
$\N ij0 = \del i{\co j}$ and
$\N ijk = \N {\co i}kj = \N k{\co j}i $ (so that it extends
to an anti-automorphism of $V$),
\item (Positive Integrality) the structure constants are
non-negative integers, $\N ijk \in \bbN_0$,
\end{enumerate}
then $(V,\cV)$ (or simply $V$) is called a sector algebra.
If $V$ is commutative, $\N ijk=\N jik$, then it is called
a fusion algebra.
\lablth{fsalg}
\end{definition}

Now let $M$ be an infinite factor and
$\cV=\{\ls 0,\ls 1, \ls 2,\ldots,\ls {d-1}\}$ be a finite
set of irreducible sectors with finite statistical
dimension, which contains the trivial
sector, say $\ls 0 = [\id]$, and is closed under
sector conjugation and the sector product. The latter
means that the irreducible decomposition of each product
$\ls i \times \ls j$ is a sum of elements in $\cV$
(possibly with some multiplicities). We simply call
such a set a {\em sector basis}. We can consider a
sector basis as the basis of an algebra $V$ where the
summation $\oplus$ and multiplication $\times$ comes
from the sum and product of sectors in the obvious
sense. By the properties of addition and multiplication
of sectors, $V$ is indeed a sector algebra, and the
structure constants are given by
$\N ijk=\la \lambda_i\circ\lambda_j,\lambda_k\ra_M$,
where $\lambda_i$ denote representative endomorphisms
of the sector $\ls i$.

Now suppose that we have a net of subfactors $\cN\subset\cM$
as described at the beginning of Section \ref{aind}.
We denote by $\LTSN\subset\Sect(\NIo)$ the set of DHR sectors,
i.e.\ the quotient of
$\DelNIo$ by inner equivalence in $\NIo$ (and similarly
$\LTSMO\subset\Sect(\MIo)$ as the quotient of $\DelMIoO$
by inner equivalence in $\MIo$). Suppose we
have a given sector basis $\cW\subset\LTSN$.  Because of
the commutativity of sectors in $\LTSN$, the associated
sector algebra $W$ is indeed a fusion algebra. As
$\alpha$-induction preserves unitary equivalence by
Corollary \ref{secpres}, the
map $\lambda\mapsto\ala$ extends to a map
$[\alpha]$: $[\lambda]\mapsto[\ala]$, from $\cW$
to $\Sect(\MIo)$. Now let $\cV$ denote the set
of all irreducible subsectors $[\beta]\in\Sect(\MIo)$
of every $[\ala]$, $[\lambda]\in\cW$. Since
$\alpha$-induction preserves the sector product and
conjugation, $\cV$ must be a sector basis and
we denote by $V$ the associated sector
algebra. However, $V$ is not necessarily
commutative. We now summarize the results of
Subsection \ref{ahom}, Lemmata \ref{mult}, \ref{sums},
\ref{conj} and Proposition \ref{abel}, in the following

\begin{theorem}
Let $\cW\subset\LTSN$
be a sector basis and $W$ the associated fusion algebra, and let
$\cV\subset\Sect(\MIo)$ be the corresponding sector basis obtained
by $\alpha$-induction and $V$ the associated sector algebra.
Then $\alpha$-induction extends to a homomorphism
$[\alpha]:W\rightarrow V$, preserving conjugates and
statistical dimensions. Each $[\ala]$,
$[\lambda]\in\cW$, commutes with each $[\beta]\in\cV$.
If $[\alpha]$ is surjective i.e.\ each element in $V$
can be written as a linear combination of
$[\alpha_{\lambda_i}]$'s, $[\lambda_i]\in\cW$, then
the sector algebra $V$ is a fusion algebra.
\lablth{IaR}
\end{theorem}

Now we turn to the discussion of $\sigma$-restriction in terms
of sectors. By Lemma \ref{sihom}, the map
$\beta\mapsto\sib$ extends to a map from $\Sect(\MIo)$ to
$\Sect(\NIo)$. We can therefore summarize the results of
Theorem \ref{asrep} and Corollary \ref{markabel} as follows.

\begin{theorem}
Let $\cT\subset\LTSMO$ be a sector basis and $T$ the
associated fusion algebra. Let also $\cW\subset\LTSN$ be
a sector basis with associated fusion algebra $W$,
and $\cV$, $V$ obtained by $\alpha$-induction as above. If
all elements of $\cT$ are mapped to elements in $W$
by $\sigma$-restriction, then $\cT\subset\cV$ and $T\subset V$ is
a (sector) subalgebra. Moreover, any element of $\cT$ commutes
with every element of $\cV$.
\lablth{mark}
\end{theorem}

\subsection{The subgroup net of subfactors}
\lablsec{subgrp}

Although we postpone all our (conformal field theory)
applications to the forthcoming
paper \cite{boev2} let us briefly discuss a simple example here.
Consider a situation as in the DHR theory \cite{dhr1}, i.e.\ we have a net
$\cF$ of local field algebras $F(I)$, $I\in\Jz$, that are
type III-factors, and we have a compact gauge group $\gfG$ acting
outerly on each $F(I)$, and this action is implemented on the
Hilbert space $\cH$ by a unitary representation $U$.
The net $\cN$ of observable algebras is then
given by the fixed point algebras $N(I)=F(I)^\gfG$. (There are also
some more physically motivated assumptions, e.g.\ certain space-time
transformation properties and that observables and fields associated
to relatively spacelike regions commute.)
Now suppose that we are dealing with a finite gauge group,
and that $\gfH\subset\gfG$ is a
subgroup. We define another net $\cM$ by taking the fixed
point algebras with respect to the subgroup, $M(I)=F(I)^\gfH$.
Then we clearly obtain a net of subfactors $\cN\subset\cM$
of finite index. (The index is in fact $[\gfG:\gfH]$.)
Under the standard assumptions of the DHR theory \cite{dhr1} the
Hilbert space $\cH$ decomposes with respect to the action of $\cN$ as
\be
\cH = \bigoplus_{\pi\in\Gh} \cH_\pi \otimes \bbC^{d_\pi} \,.
\labl{decomG}
Here $\pi\in\Gh$ are the irreducible representations of $\gfG$ of
dimension $d_\pi$, and $\cH_\pi$ are pairwise inequivalent
representation spaces of $\cN$, the superselection sectors.
The gauge group $\gfG$ acts on the multiplicity spaces
$\bbC^{d_\pi}$ by the representation $\pi$, i.e.\
\be
U(g)=\bigoplus_{\pi\in\Gh} \bfe_{\cH_\pi} \otimes \pi(g)\,,
\qquad g\in\gfG \,.
\labl{gauge}
With respect to $\cM$ we have another decomposition
\be
\cH = \bigoplus_{\rho\in\Hh} \cH_\rho \otimes \bbC^{d_\rho} \,,
\labl{decomH}
where now $\rho\in\Hh$ label the irreducible representations
(of dimension $d_\rho$) of the subgroup $\gfH$. Since
$N(I)=F(I)\cap U(\gfG)'$ and $M(I)=F(I)\cap U(\gfH)'$ it is not
hard to see that the decompositions of \erf{decomG} and
\erf{decomH} are related by
\be
\cH_\rho = \bigoplus_{\pi\in\Gh} \cH_\pi \otimes
\bbC^{n^\pi_\rho} \,,
\labl{relGH}
where $n^\pi_\rho$ are the induction-restriction coefficients
\[ n^\pi_\rho = \la \rho, \res \gfG\gfH \pi \ra_{\bbZ[\Hh]}
= \la \ind \gfG\gfH \rho, \pi \ra_{\bbZ[\Gh]} \,. \]
Now let us assume that our requirements of Haag duality and
strong additivity for the net $\cN$ and locality of the
net $\cM$ are fulfilled. Let
$\lambda_\pi\in\DelNIo$ and $\beta_\rho\in\DelMIoO$,
$\Io\in\Jz$, denote localized endomorphisms corresponding to the
superselection sectors $\cH_\pi$, $\pi\in\Gh$, and $\cH_\rho$,
$\rho\in\Hh$, so that they obey in particular the
fusion rules of $\Gh$ and $\Hh$, respectively, and their
statistical dimensions coincide with the dimensions of the
corresponding group representations. We learn from Proposition
\ref{LR2} (see also \cite{lore}) that $\sigma$-restriction
corresponds to the restriction of representations of the net $\cM$
to the net $\cN$, i.e.\
$\pi^0\circ\beta_\rho|_{\cN}\simeq\pio\circ\sigma_{\beta_\rho}$.
This restriction can be read off from \erf{relGH}, hence we conclude
for $\rho\in\Hh$
\[ [\sigma_{\beta_\rho}] = \bigoplus_{\pi\in\Gh} n^\pi_\rho \,
[\lambda_\pi] \,. \]
>From $\alpha\sigma$-reciprocity, Theorem \ref{asrep},
\[ \la \alpha_{\lambda_\pi},\beta_\rho \ra_\MIo = \la
\lambda_\pi, \sigma_{\beta_\rho} \ra_\NIo = n^\pi_\rho \,, \]
we conclude (recall $d_{\ala}=d_\lambda$)
\[ [\alpha_{\lambda_\pi}] = \bigoplus_{\rho\in\Hh} n^\pi_\rho \,
[\beta_\rho] \,. \]
In other words, for this particular example of
the subgroup net of subfactors,
$\sigma$-restriction corresponds to the induction,
$\alpha$-induction corresponds to the restriction of
group representations, and $\alpha\sigma$-reciprocity
reflects Frobenius reciprocity.

\subsection{Remarks}

In view of our later applications to chiral conformal
field theories \cite{boev2} we
have presented the theory for nets of subfactors indexed by the
set $\Jz$, i.e.\ with the punctured circle $S^1\setminus \{z\}$
as the underlying ``space-time'', and we also required strong
additivity of $\cN$ or, equivalently, of $\cA$. Note that for
chiral Conformal field theories strong additivity is equivalent to the already
assumed Haag duality (on the punctured circle). For the
general case we assumed strong additivity so that local
intertwiners (of localized endomorphisms) extend to global
ones and therefore satisfy the naturality equations and BFEs.
One may however drop the strong additivity assumption and work
with global intertwiners from the beginning. The invariance of
local algebras $M(I)$, $I\in\Jz$, $\Io\subset I$,
under the action of $\ala$ is also true without the strong
additivity assumption because $v$ itself is a global intertwiner.
Moreover, many of our results possess global analogues,
e.g.\ Theorem \ref{main} then reads
$\la\ala,\amu\ra_\cM=\la\canr\circ\lambda,\mu\ra_\cN$ for
$\lambda,\mu\in\DelNIo$ or Theorem \ref{asrep} becomes
$\la\ala,\beta\ra_\cM=\la\lambda,\sib\ra_\cN$, $\beta\in\DelMIoO$.
However, we cannot obtain Corollary \ref{surp} without the
strong additivity assumption and we need the local formulation
for the results concerning the subsectors of the $[\ala]'s$.
But global analogues of the results not depending on the
strong additivity can also be generalized
to other space-times like the $D$-dimensional Minkowski space
$\bbM^D$ with $D=2,3,4,...$ (as long as we have transportable
endomorphisms).  One just has to replace intervals $I$ by
double cones $\cO$ and to substitute ``disjoint'',
$I_1\cap I_2=\emptyset$, by ``causally disjoint'', i.e.\
``relatively spacelike'', $\cO_1\subset\cO_2'$. But notice
that for $\bbM^D$ with $D>2$ the spacelike complement of any
double cone is connected, and this implies that we only have one
statistics operator. There are no longer two different braidings
and hence we have $\ala\equiv\ala^+=\ala^-$, and in particular
all induced endomorphisms $\ala$ are localized.

\subsection*{Acknowledgement}

We are grateful to K.-H.\ Rehren for several useful comments
on an earlier version of the manuscript.
This project is supported by the EU TMR Network in
Non-Commutative Geometry.



\newcommand\biba[7]   {\bibitem{#1} {\sc #2:} {\sl #3.} {\rm #4} {\bf #5}
                      { (#6) } {#7} }
\newcommand\bibb[4]   {\bibitem{#1} {\sc #2:} {\it #3.} {\rm #4}
                      } 
\newcommand\bibp[4]   {\bibitem{#1} {\sc #2:} {\sl #3.} {\rm Preprint #4}
                      } 
\newcommand\bibx[4]   {\bibitem{#1} {\sc #2:} {\sl #3} {\rm #4}
                      } 
\def\AAM              {Acta Appl.\ Math.}
\def\CMP              {Com\-mun.\ Math.\ Phys.}
\def\IJM              {Intern.\ J. Math.}
\def\JFA              {J.\ Funct.\ Anal.}
\def\JMP              {J.\ Math.\ Phys.}
\def\LMP              {Lett.\ Math.\ Phys.}
\def\RMP              {Rev.\ Math.\ Phys.}
\def\Inv              {Invent.\ Math.}
\def\npbp             {Nucl.\ Phys.\ {\bf B} (Proc.\ Suppl.)}
\def\nupb             {Nucl.\ Phys.\ {\bf B}}
\def\adma             {Adv.\ Math.}
\def\coma             {Con\-temp.\ Math.}
\def\physa            {Physica {\bf A}}
\def\ijmp             {Int.\ J.\ Mod.\ Phys.\ {\bf A}}
\def\FdP              {Fortschr.\ Phys.}
\def\PLB              {Phys.\ Lett.\ {\bf B}}
\def\RIMS             {Publ.\ RIMS, Kyoto Univ.}



\end{document}